\documentclass[letterpaper,10pt]{article}

\usepackage[utf8]{inputenc}
\usepackage[left=2cm, right=2cm, top=2cm, bottom=2cm]{geometry}
\usepackage[english]{babel}
\usepackage{amsmath,empheq}
\usepackage{amsfonts}
\usepackage{amssymb}
\usepackage{graphicx}
\usepackage{url}
\usepackage{caption}
\usepackage{keyval}
\usepackage{etex}
\usepackage{subfig}
\usepackage{placeins}
\usepackage{authblk}

\usepackage{amsthm}

\usepackage{xcolor}
\definecolor{verde}{rgb}{0,0.6,0}
\definecolor{darkgreen}{rgb}{0,0.3,0}

\newtheorem{proposition}{Proposition}

\title{A SIDARTHE Model of COVID-19 Epidemic in Italy}

\author[1]{Giulia Giordano}
\author[2]{Franco Blanchini}
\author[3,4]{Raffaele Bruno}
\author[5]{Patrizio Colaneri}
\author[3]{Alessandro Di Filippo}
\author[3]{Angela Di Matteo}
\author[3]{Marta Colaneri}
\author[6]{the COVID19 IRCCS San Matteo Pavia Task Force.}

\affil[1]{Department of Industrial Engineering, University of Trento, Italy. {\tt giulia.giordano@unitn.it}} 
\affil[2]{Dipartimento di Scienze Matematiche, Informatiche e Fisiche, University of Udine, Italy. {\tt blanchini@uniud.it}}
\affil[3]{Division of Infectious Diseases I, Fondazione IRCCS Policlinico San Matteo, Pavia, Italy. {\tt raffaele.bruno@unipv.it, alessandro.difilippo01@universitadipavia.it, a.dimatteo@smatteo.pv.it, marta.colaneri@universitadipavia.it}}
\affil[4]{Department of Clinical, Surgical, Diagnostic, and Paediatric Sciences, University of Pavia, Italy.}
\affil[5]{Dipartimento di Elettronica, Informazione e Bioingegneria, Politecnico di Milano, and IEIIT-CNR, Italy. {\tt patrizio.colaneri@polimi.it}}
\affil[6]{Fondazione IRCCS Policlinico San Matteo, Pavia, Italy.}

\date{March 21, 2020}

\begin{document}
\maketitle

\begin{abstract} 
In late December 2019, a novel strand of Coronavirus (SARS-CoV-2) causing a severe, potentially fatal respiratory syndrome (COVID-19) was identified in Wuhan, Hubei Province, China and is causing outbreaks in multiple world countries, soon becoming a pandemic. 
Italy has now become the most hit country outside of Asia: on March 16, 2020, the Italian Civil Protection documented a total of 27980 confirmed cases and 2158 deaths of people tested positive for SARS-CoV-2.
In the context of an emerging infectious disease outbreak, it is of paramount importance to predict the trend of the epidemic in order to plan an effective control strategy and to determine its impact.
This paper proposes a new epidemic model that discriminates between infected individuals depending on whether they have been diagnosed and on the severity of their symptoms.
The distinction between diagnosed and non-diagnosed is important because non-diagnosed individuals are more likely to spread the infection than diagnosed ones, since the latter are typically isolated, and can explain misperceptions of the case fatality rate and of the seriousness of the epidemic phenomenon. Being able to predict the amount of patients that will develop life-threatening symptoms is important since the disease frequently requires hospitalisation (and even Intensive Care Unit admission) and challenges the healthcare system capacity.
We show how the basic reproduction number can be redefined in the new framework, thus capturing the potential for epidemic containment. 
Simulation results are compared with real data on the COVID-19 epidemic in Italy, to show the validity of the model and compare different possible predicted scenarios depending on the adopted countermeasures.
\end{abstract}


\section{Introduction}

In late December 2019, a novel strand of Coronavirus (SARS-CoV-2) causing a severe, potentially fatal respiratory syndrome (COVID-19) was identified in Wuhan, Hubei Province, China \cite{Velavan2020,Wu2020}. Starting from December 18, 2019, an exponentially growing number of patients in mainland China were admitted to hospitals with a diagnosis of COVID-19, prompting Chinese authorities to introduce radical measures to contain the outbreak \cite{Guan2020}.

Despite those efforts, the virus had managed to spread to several parts of the world: a report from WHO dated March 4, 2020 documented a grand total of 93090 confirmed cases spanning 77 different countries worldwide \cite{WHO2020}. Italy has since become the most hit country outside of Asia \cite{2Remuzzi}.
After the first indigenous case was confirmed on February 21, 2020, in the Lodi province (Lombardy, northern Italy), several suspect cases, initially epidemiologically linked, began to emerge in the south and southwest territory of Lombardy \cite{Giuffrida}.

A \emph{red zone}, in which SARS-CoV-2 infection is considered endemic, was instituted on February 22, 2020, encompassing 11 municipalities all surrounding and including the epicenter of the infection; those cities were put on lockdown to try and contain the emerging threat.
Regional health authorities promoted an aggressive campaign to identify and screen all close contacts with confirmed cases of COVID-19 that resulted, up until March 16, 2020, in the execution of 137962 nasal swabs for the detection of the virus. Of the overall 27980 detected cases in Italy at the time of the report, 23073 were currently infected (particularly, 11025 were hospitalised, 1851 were admitted to Intensive Care Units, 10197 were quarantined at home), 2749 were considered clinically healed and discharged and 2158 died \cite{MSalute}.
In the first days of the outbreak, both symptomatic and asymptomatic subjects underwent screening, after which a Government regulation dated February 26, 2020 limited screening to symptomatic subjects only, due partly to a shortage of material and partly to most tests resulting negative, according to Italian Civil Protection data \cite{PC,Locatelli}.
On March 8, 2020, in an effort to further contain the spread of SARS-CoV-2, an administrative act extended the \emph{red zone} to the entire area of Lombardy and 14 more northern Italian provinces. Eventually, on the evening of March 9, 2020, the entire country was declared as \emph{red zone} \cite{DPCM}.

COVID-19 displays peculiar epidemiological traits when compared to previous outbreaks due to related viruses such as SARS-CoV and MERS-CoV.
According to early Chinese data \cite{Wang2020}, a large quota of transmissions, both in a nosocomial and in a community setting, occurred through human-to-human contacts with individuals showing no or mild symptoms of the disease.
The estimated basic reproduction number ($R_0$) for SARS-CoV-2 is higher than both for SARS-CoV and for MERS-CoV (ranging from 2.0 to 3.5) \cite{Fisman2004,Zhao2020,Read2020}.
Additionally, high viral loads of SARS-CoV-2 were found in upper respiratory specimens of patients showing little or no symptoms, with a viral shedding pattern akin to that of influenza viruses \cite{Zou2020}. These findings suggest that inapparent transmission may play a major and underestimated role in sustaining the outbreak.

In the context of an emerging infectious disease outbreak, predicting the trend of the epidemic is of paramount importance to plan effective control strategies and determine how said strategies impact the course of the epidemic. Predictive mathematical models have been formulated over the years \cite{AndersonMay1991,DiekmannHeesterbeek2000,Hethcote2000,BrauerCastilloChavez2012}, including the SIR model, a simple and widely-used deterministic model for human-to-human transmission that describes the flow of individuals through three mutually exclusive stages of infection: Susceptible (S), Infected (I) and Recovered (R) \cite{Kermack1927}.

However, more complex models are needed to accurately portray the dynamic characteristics of specific epidemics. To this end, we propose a new mean-field epidemiological model, inspired by the COVID-19 epidemic in Italy, that serves as an extension of the classical SIR model in the wake of Gumel et al., who developed a variant dynamic mathematical model that reflected some key epidemiological properties of SARS \cite{Gumel2004}. For COVID-19, Lin et al. considered an extended SEIR model including a variable representing risk perception and a variable representing the cumulative number of cases \cite{QLin2020}, while Anastassopoulou et al. proposed a discrete-time SIR model including a variable for Dead individuals \cite{Anastassopoulou2020}, and Casella proposed a control-oriented SIR model stressing the role of delays, to compare the outcome of different policies aimed at containing the epidemics \cite{Casella2020}, while Wu et al. observed transmission dynamics to estimate clinical severity \cite{JWu2020}; stochastic transmission models have been also considered, see for instance \cite{Hellewell2020,Kucharski2020}.

Our new model, named SIDARTHE, discriminates between detected and undetected cases of infection, either asymptomatic or symptomatic, and also between different Severity of Illness (SOI): non-life-threatening cases (minor and moderate infection) and potentially life-threatening cases (major and extreme) that require Intensive Care Unit (ICU) admission. 
We omit the probability rate of becoming susceptible again, after having already recovered from the infection. Although anecdotal cases  are found in literature, the reinfection rate value appears to be negligible based on early evidence \cite{Lan2020}.

We estimate the model parameters based on national data about the evolution of the epidemic in Italy in the period from February 20, 2020 to March 12, 2020. Based on this model calibration, we discuss possible longer-term scenarios that showcase the impact of different countermeasures to contain the contagion.

\section{Results}

We present here the proposed model and we analyse it: we study possible equilibria and their stability, which are related to the possible long-term outcomes of the epidemics, and we generalise to our model the concept of the basic reproduction number, $R_0$, which is the expected number of new cases directly generated by a single infected subject; in SIR models, the basic reproduction number corresponds to the ratio between the probability of contact between Susceptible and Infected, leading to contagion, and the probability of turning from Infected to Recovered, under the ideal assumption that all actual cases of infection are recognised as such.

Then, using the available official data to estimate the model parameters, we study the evolution of the COVID-19 epidemic in Italy, comparing possible scenarios resulting from different social-distancing measures.

\subsection{SIDARTHE Mathematical Model}

\begin{figure}[h!]
\centering
\vspace{5mm}
\includegraphics[width=10cm]{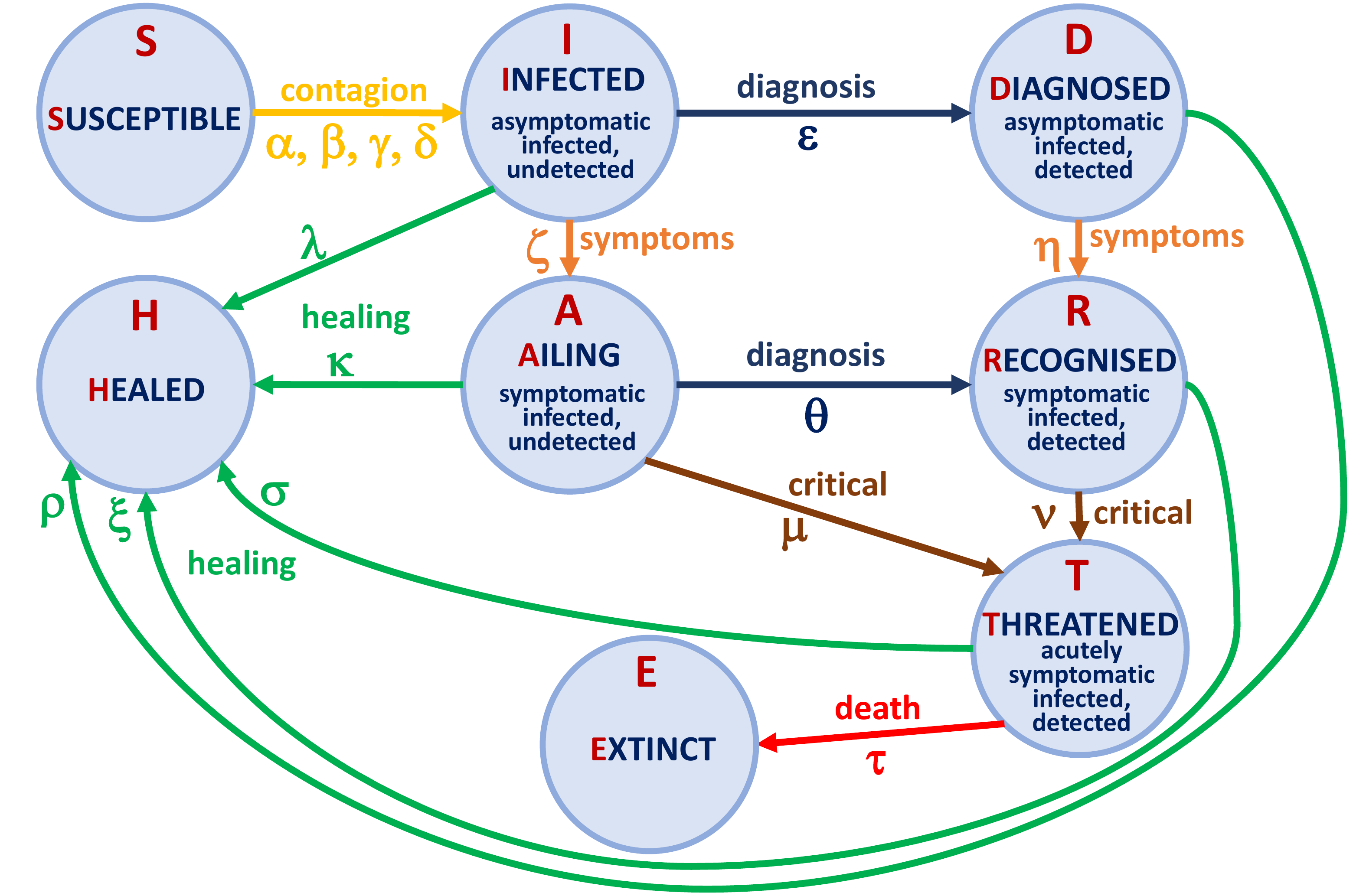}
\caption{Graphical scheme representing the interactions among different stages of infection in the mathematical model SIDARTHE.}
\label{SIDHART_scheme}
\vspace{5mm}
\end{figure}

In our model, the total population is partitioned into the following eight stages: S, Susceptible; I, Infected (asymptomatic infected, undetected); D, Diagnosed (asymptomatic infected, detected); A, Ailing (symptomatic infected, undetected); R, Recognised (symptomatic infected, detected);
T, Threatened (infected with life-threatening symptoms, detected); H, Healed (recovered); E, Extinct (dead).
The interactions among these eight stages and the related interaction parameters are shown in Figure \ref{SIDHART_scheme}.

The SIDARTHE dynamical system consists of eight ordinary differential equations, describing the evolution of the population in each stage over time:
\begin{eqnarray}
\dot S(t) &=&-S(t)(\alpha I(t)+\beta D(t)+\gamma A(t)+\delta R(t)) \label{eq:S}\\
\dot I(t) &=&S(t)(\alpha I(t)+\beta D(t)+\gamma A(t)+\delta R(t))- (\epsilon+\zeta+\lambda)I(t)\label{eq:I}\\
\dot D(t)&=&\epsilon I(t)-(\eta+\rho)D(t) \label{eq:D}\\
\dot A(t)&=& \zeta I(t)-(\theta+\mu+\kappa)A(t) \label{eq:A}\\
\dot R(t)&=& \eta D(t)+\theta A(t)-(\nu+\xi)R(t) \label{eq:R}\\
\dot T(t)&=& \mu A(t)+\nu R(t)-(\sigma+\tau)T(t) \label{eq:T}\\
\dot H(t)&=&\lambda I(t)+\rho D(t)+\kappa A(t)+\xi R(t)+\sigma T(t) \label{eq:H}\\
\dot E(t)&=& \tau T(t) \label{eq:E}
\end{eqnarray}
where the uppercase Latin letters (state variables) represent the fraction of population in each stage, while all the considered parameters, denoted by Greek letters, are positive numbers.
In particular, the parameters:
\begin{itemize}
\item $\alpha$, $\beta$, $\gamma$, $\delta$ respectively denote the transmission rate (i.e. the probability of disease transmission in a single contact times the average number of contacts per person) due to contacts between a Susceptible subject and an Infected, a Diagnosed, an Ailing, a Recognised subject. Typically, $\alpha$ is larger than $\gamma$ (assuming that people tend to avoid contacts with subjects showing symptoms, even though diagnosis has not been made yet), which in turn is probably larger than $\beta$ and $\delta$ (assuming that subjects who have been diagnosed are properly isolated). These parameters can be modified by social distancing policies (e.g., closing schools, remote working, etc.). The risk of contagion due to Threatened subjects, treated in proper ICUs, is assumed negligible.
\item $\epsilon$ and $\theta$ capture the probability rate of detection, relative to asymptomatic and mildly symptomatic cases respectively. These parameters, also modifiable, reflect the level of attention on the disease and the number of tests performed over the population. Note that $\theta$ is typically larger than $\epsilon$, since a symptomatic individual is more likely to get tested.
\item $\zeta$ and $\eta$ denote the probability rate at which an infected subject, respectively not aware and aware of being infected, develops clinically relevant symptoms, and are probably comparable. These parameters are disease-dependent and hardly modifiable.
\item $\mu$ and $\nu$ respectively denote the rate at which undetected and detected infected subjects develop life-threatening symptoms, and are likely to be comparable if there is no known specific treatment that is effective against the disease, otherwise $\mu$ is likely to be larger. These parameters can be reduced by means of improved therapies and acquisition of immunity against the virus.
\item $\tau$ denotes the mortality rate (for infected subjects with life-threatening symptoms) and can be reduced by means of improved therapies.
\item $\lambda$, $\kappa$, $\xi$, $\rho$ and $\sigma$ denote the rate of recovery for the five classes of infected subjects, and may differ significantly if an appropriate treatment for the disease is known and adopted to diagnosed patients, while are probably comparable otherwise. These parameters can be increased thanks to improved treatments and acquisition of immunity against the virus.
\end{itemize}

We omit the probability rate of becoming susceptible again, after having already recovered from the infection, since its value appears to be negligible (although not zero) based on early evidence \cite{Lan2020,Bai2020}.

\subsection{Analysis of the Mathematical Model}

The SIDARTHE model \eqref{eq:S}-\eqref{eq:E} is a bilinear system with $8$ differential equations. The system is positive: all the state variables take nonnegative values for $ t\ge 0$ if initialised at time $0$ with nonnegative values. 
Note that $H(t)$ and $E(t)$ are cumulative variables that depend only on the other ones and their own initial conditions.

The system is compartmental and has the mass conservation property: as it can be immediately checked, $\dot S(t)+\dot I(t)+\dot D(t)+\dot A(t)+\dot R(t)+\dot T(t)+\dot H(t)+\dot  E(t)=0$, hence the sum of the states (total population) is constant.
Since the variables denote population \emph{fractions}, we can assume:
$$S(t)+I(t)+ D(t)+ A(t)+R(t)+T(t)+ H(t) + E(t)=1,$$
where $1$ denotes the total population, including deceased.

Given an initial condition
$S(0)$, $I(0)$, $D(0)$, $A(0)$, $R(0)$, $T(0)$, $H(0)$, $E(0)$ summing up to $1$,
the variables   converge to an equilibrium
$$ \bar S \geq 0 ,~~ \bar I=0, ~\bar D =0,~\bar A=0, ~\bar R = 0, ~\bar T = 0,~~\bar H\geq 0,~~\bar E\geq 0,$$
with $\bar S+\bar H + \bar E=1$, as shown in the Methods section. So only the susceptible, the healed and the deceased
populations are eventually present, meaning that the epidemic phenomenon is over.

To understand the system behaviour, we partition it into three subsystems: the first including just variable $S$ (corresponding to susceptible individuals), the second including variables $H$ and $E$ (representing healed and defuncts), and the third including $I$, $D$, $A$, $R$ and $T$ (the infected individuals), which are nonzero only during the transient. We focus on the third subsystem, which we denote as the $IDART$ subsystem. An important observation is that when (and only when) the infected individuals $I+D+A+R+T$ are zero, the remaining variables $S$, $H$ and $E$ are at the equilibrium. Variables $H$ and $E$ (which are monotonically increasing) converge to their asymptotic values $\bar H$ and $\bar E$, and $S$ (which is monotonically decreasing) converges to $\bar S$ if and only if $I$, $D$, $A$, $R$ and $T$ converge to zero.

\begin{figure}[t]
\centering
\includegraphics[width=5cm]{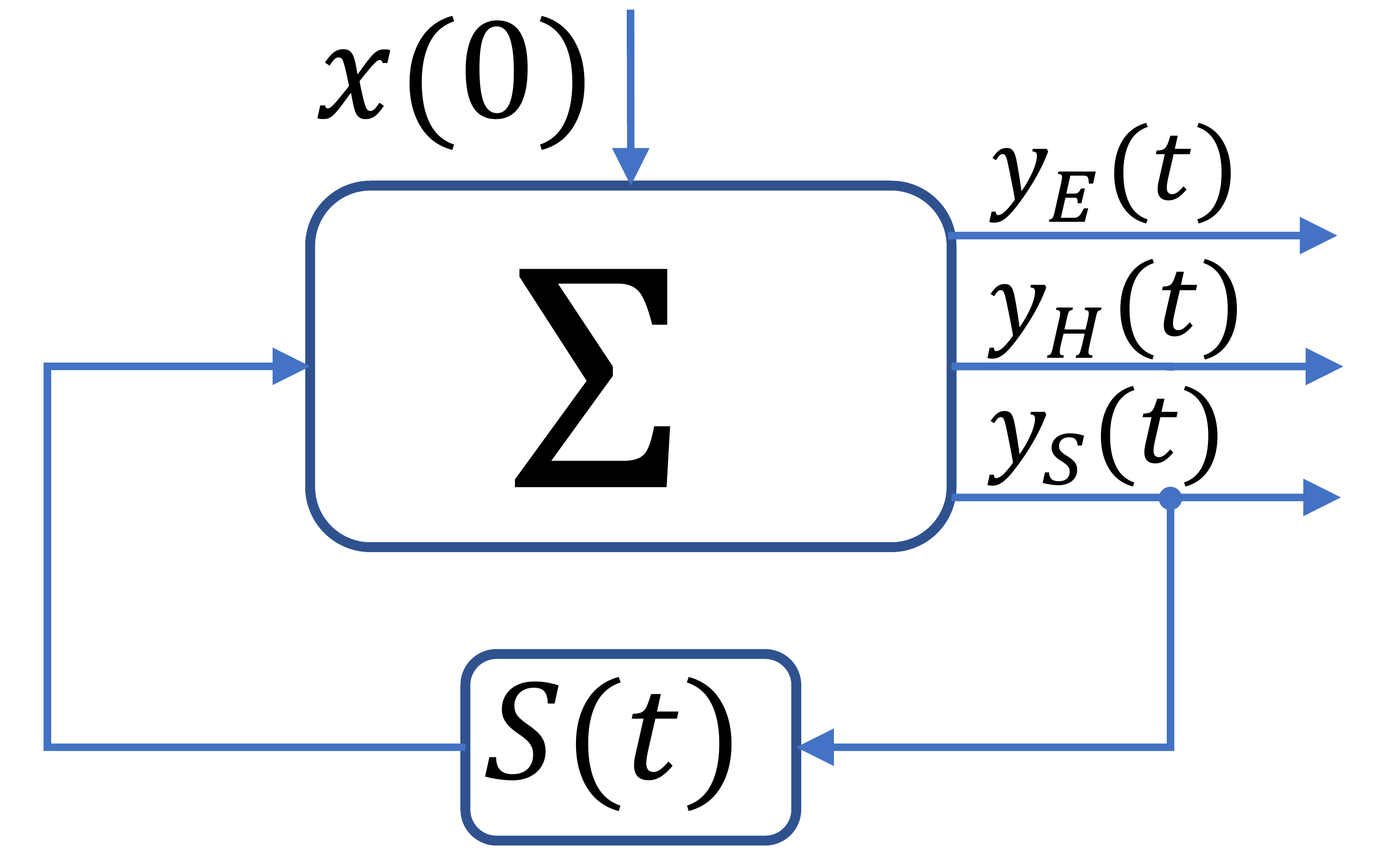}
\caption{Intrinsic feedback in the SIDARTHE model: the $IDART$ subsystem, $\Sigma$, can be seen as a positive linear system under feedback, and variables $S$, $H$ and $E$ can be computed based on its outputs.}
\label{Fig_intrinsic_feedback}
\end{figure}

The overall system can be recast in the feedback structure shown in Figure \ref{Fig_intrinsic_feedback}: the $IDART$ subsystem can be seen as a positive linear systems subject to feedback.

To be precise, defining  $x=[I \, D \, A \, R \, T]^\top$, we can rewrite the $IDART$ subsystem as 
\begin{eqnarray}
\dot x(t)&=&Fx(t)+bu(t)=\begin{bmatrix} -r_1 & 0 & 0 & 0 & 0\\  \epsilon & -r_2 & 0 & 0 & 0\\  \zeta & 0 & -r_3& 0 & 0\\ 0&  \eta &\theta &-r_4 & 0\\ 0 & 0 & \mu & \nu & -r_5\end{bmatrix} x(t)+\begin{bmatrix} 1 \\ 0 \\ 0 \\ 0\\0\end{bmatrix}u(t) \label{eq:IDAR}\\
y_S(t)&=& c^\top x(t)=\begin{bmatrix} \alpha & \beta &  \gamma &  \delta & 0\end{bmatrix}x(t) \label{eq:yS}\\
y_H(t)&=& f^\top x(t) =\begin{bmatrix} \lambda & \rho &  \kappa &  \xi & \sigma\end{bmatrix}x(t) \label{eq:yH}\\
y_E(t)&=& d^\top x(t)= \begin{bmatrix} 0& 0 &  0 &  0 & \tau\end{bmatrix}x(t)\label{eq:yT}\\
u(t)&=& S(t)y_S(t)\label{eq:u}
\end{eqnarray}
where $r_1=\epsilon+\zeta+\lambda$, $r_2=\eta+\rho$, $r_3=\theta+\mu+\kappa$, $r_4=\nu+\xi$, $r_5 = \sigma+\tau$. 
The remaining variables satisfy the differential equations
\begin{eqnarray}
\dot S(t)&=&-S(t)y_S(t) \label{eq:S_y}\\
\dot H(t)&=&y_H(t) \label{eq:H_y}\\
\dot E(t)&=&y_E(t) \label{eq:E_y}
\end{eqnarray}
Since the time varying feedback gain $S(t)$ eventually converges to a constant value $\bar S$,
we can proceed with a parametric study with respect to the asymptotic feedback gain $\bar S$. A fundamental result is given in the following proposition, proved in the Methods section.
\begin{proposition} \label{stab}
The $IDART$ subsystem with constant feedback $\bar S$ is asymptotically stable if and only if
 \begin{equation}\label{threshold}
\bar S<\bar S^*=\frac{r_1r_2r_3r_4}{\alpha r_2r_3r_4+\beta\epsilon r_3r_4+\gamma\zeta r_2r_4+\delta(\eta\epsilon r_3+\zeta\theta r_2)}.
\end{equation}
\end{proposition} 
The threshold $\bar S^*$ is of fundamental importance. 
Since asymptotically $S(t)$ converges monotonically to a constant $\bar S$, such a constant $\bar S$ must ensure convergence of the $IDART$ subsystem to zero, hence stability (otherwise, $S$ could not converge to $\bar S$). As a consequence, we have the following result.
\begin{proposition} \label{pro:threshold}
For positive initial conditions, the limit value $\bar S = \lim_{t \rightarrow \infty} S(t)$
cannot exceed $\bar S^*$.
\end{proposition} 
The proof is in the Methods section.

The threshold value \eqref{threshold} has a deep meaning. The limit $\bar S$ represents the fraction of population that has never been infected.
This value is a decreasing function of the parameters $\alpha$, $\beta$, $\gamma$ and $\delta$, which are the infection parameters. The feedback action
$$
u(t) = S(t) y_S(t) = S(t) ( \alpha I + \beta D + \gamma A + \delta D)
$$
has a destabilizing effect on the $IDART$ subsystem, which would be stable without the feedback.
To preserve stability of the $IDART$ subsystem and ensure that the equilibrium $\bar S$ is reached, either the infection coefficients are small or the final value $\bar S$ is small.
Defining the basic reproduction number as 
\begin{equation}\label{eq.R0}
R_0:=\frac{1}{\bar S^*}=
\frac{\alpha}{r_1} +  \frac{\beta\epsilon}{r_1r_2}+
\frac{\gamma\zeta}{r_1r_3}+\frac{\delta\eta\epsilon}{r_1 r_2r_4}+\frac{\delta \zeta\theta}{r_1r_3r_4},
\end{equation} 
we have that stability of the equilibrium occurs for 
\begin{equation}\label{threshold_R0}
\bar S\bar R_0<1.
\end{equation}
At the outset of the epidemic we have $\bar S\simeq 1$, so that stability occurs for 
$$
R_0<1,
$$
which essentially represents an immediate recover with no large involvement of the population.
Larger values of $R_0$ imply a strong affection of the population according to \eqref{threshold_R0}.

We can provide an important formula that relates the coefficient $R_0$ with the steady-state
value $\bar S$ (and $\bar H$, $\bar E$). The derivation is in the Methods section.  
\begin{proposition} \label{pro:steady}
For positive initial conditions, the limit values of  $\bar S = \lim_{t \rightarrow \infty} S(t)$, $\bar H = \lim_{t \rightarrow \infty} H(t)$ and $\bar E = \lim_{t \rightarrow \infty} E(t)$ are given by 

\begin{equation} \label{R0S}
f_0 + R_0 (S(0) - \bar S) = \log \left ( \frac{S(0)}{\bar S} \right )
\end{equation}

\begin{equation} \label{R0H}
\bar H=H(0)+f_H + R_H (S(0) - \bar S)
\end{equation}

\begin{equation} \label{R0E}
\bar E=E(0)+f_E + R_E (S(0) - \bar S)
\end{equation}

where $f_0= -c^\top F^{-1}x(0)$, $f_H= -f^\top F^{-1}x(0)$, $f_E= -d^\top F^{-1}x(0)$, $R_H = -f^\top F^{-1} b$ and $R_E = -d^\top F^{-1} b$.  

\end{proposition} 

If we consider an initial condition in which only \emph{undiagnosed} infected $I(0)>0$ are present, while $D(0)=A(0)=R(0)=0$, then we can explicitly compute $f_0= -c^\top F^{-1}\begin{bmatrix} I(0)& 0& 0& 0 & 0\end{bmatrix}^\top$ as
\begin{equation} \label{f0} 
f_0 = R_0 I(0).
 \end{equation}
It is important to stress that \eqref{f0} could be totally misleading for a long term prediction, because in the long run the coefficients of matrix $F$ are going to change.
So, if there is a change in the parameters at a time $t_0$, for instance due to imposed restrictions and countermeasures, the prediction has to be adjusted by considering $f_0=- c^\top F^{-1}x(t_0)$, where $x(t_0)=[I(t_0)~D(t_0)~A(t_0)~R(t_0)~T(t_0)]^\top$ and $F$ includes the new parameter values.
Clearly also \eqref{R0S} has to be updated by considering the new $S(t_0)$.

An important indicator of the dynamics of epidemiologic model is the case fatality rate, CFR, which is the ratio between number of deaths and number of infected. Our model allows us to distinguish between the actual CFR $M(t)$ and the perceived CFR $P(t)$, which are defined as
\begin{eqnarray}
\label{eq.M}
M(t)&=&\frac{E(t)}{\int_0^t S(\phi)[\alpha I(\phi)+\beta D(\phi)+\gamma A(\phi)+\delta R(\phi)]d\phi}\\
\label{eq.P}
P(t)&=&\frac{E(t)}{\int_0^t [\epsilon I(\phi)+(\theta+\mu) A(\phi)]d\phi}
\end{eqnarray}

Taking into account that 
\begin{equation}
S(t)=S(0)+I(0)-I(t)-r_1\int_0^t I(\phi)d\phi,
\end{equation}
we can provide the explicit formulas
\begin{eqnarray}
\label{eq.M1}
M(t)&=&\frac{E(t)}{S(0)-S(t)}\\
\label{eq.P1}
P(t)&=&\frac{E(t)}{\frac{\epsilon r_3+(\theta+\mu)\zeta}{r_1r_3}[I(0)+S(0)-I(t)-S(t)]+\frac{\theta+\mu}{r_3}[A(0)-A(t)]}
\end{eqnarray}
with equilibria
\begin{eqnarray}
\label{eq.barM}
\bar M&=&\frac{\bar E}{S(0)-\bar S}\\
\label{eq.barP}
\bar P&=&\frac{\bar E}{\frac{\epsilon r_3+(\theta+\mu)\zeta}{r_1r_3}[I(0)+S(0)-\bar S]+\frac{\theta+\mu}{r_3}A(0)}
\end{eqnarray}

\subsection{Case Study: the COVID-19 Outbreak in Italy}

We consider the outbreak of COVID-19 epidemic in Italy. We estimate the model parameters based on data about the evolution of the epidemic in Italy in the period from February 20, 2020 to March 12, 2020; more details are in the Methods section. Based on this model calibration, we discuss possible longer-term scenarios that showcase the impact of different countermeasures to contain the contagion.

The fraction of the population in each stage at day $1$ is set as: $I=200/60e6$, $D=20/60e6$, $A=1/60e6$, $R=2/60e6$, $T=0$, $H=0$, $E=0$; $S=1-I-D-A-R-T-H-E$.
The parameters are set as
$\alpha=0.570$,
$\beta=\delta=0.011$,
$\gamma=0.456$,
$\epsilon=0.171$,
$\theta=0.371$,
$\zeta=\eta=0.125$,
$\mu=0.012$,
$\nu=0.027$,
$\tau=0.003$,
$\lambda=\rho=0.034$,
$\kappa=\xi=\sigma=0.017$.
The resulting basic reproduction number is $R_0=2.38$, which leads to a significant growth of the number of infected individuals.

After day $4$, as a consequence of basic social distancing measures due to the public opinion being aware of the epidemic outbreak and due to recommendations (such as washing hands often, not touching one's face, avoiding handshakes and keeping distance) and early measures (such as closing schools) by the Italian government, we set $\alpha=0.422$, $\gamma=0.285$ and $\beta=\delta=0.0057$, thus the new basic reproduction number becomes $R_0=1.66$.

Also, after day $12$, we set $\epsilon = 0.143$ as a consequence of the policy limiting screening to symptomatic individuals only; thus, totally asymptomatic individuals are almost no longer detected, while individuals with very mild symptoms are still detected (hence $\epsilon$ is not set exactly to zero).

Then, the comparison between the data and the curves resulting from the SIDARTHE model is reported in Figure \ref{fig:FIT}, which shows the epidemic evolution in the first $23$ days.
Figure \ref{fig:Case1} (a) shows the evolution of the full model with the estimated parameters over this short horizon and highlights that, in the earliest epidemic phase, the number of infected is significantly underestimated: almost $50\%$ of the cases are not diagnosed; Figure \ref{fig:Case1} (b) shows how the infected individuals are partitioned into the different sub-populations (diagnosed or not, with different SOI classification).

Based on the obtained model calibration, we can discuss possible longer-term scenarios that showcase the impact of different countermeasures to contain the contagion (based on a completely different model, the impact of different mitigation measures for pandemic influenza in Italy was compared in \cite{Ciofi2008}).

\subsubsection{Simulation of Intervention Scenarios}

\begin{figure}[htb]
        \centering
        \subfloat[Short-term evolution: actual vs. diagnosed cases of infection.]{\includegraphics[width=.48\textwidth]{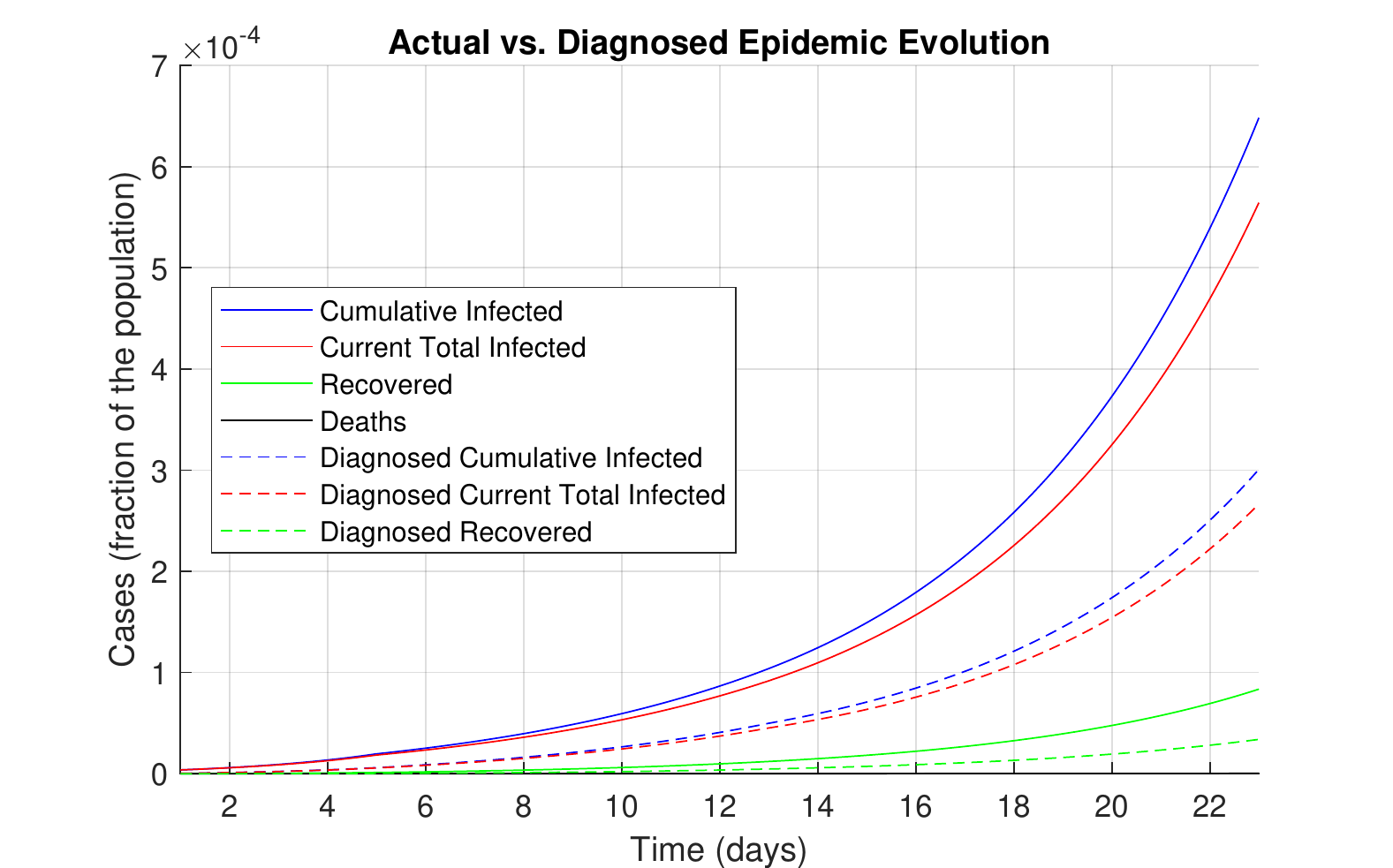}} \quad
        \subfloat[Short-term evolution of infected sub-populations.]{\includegraphics[width=.48\textwidth]{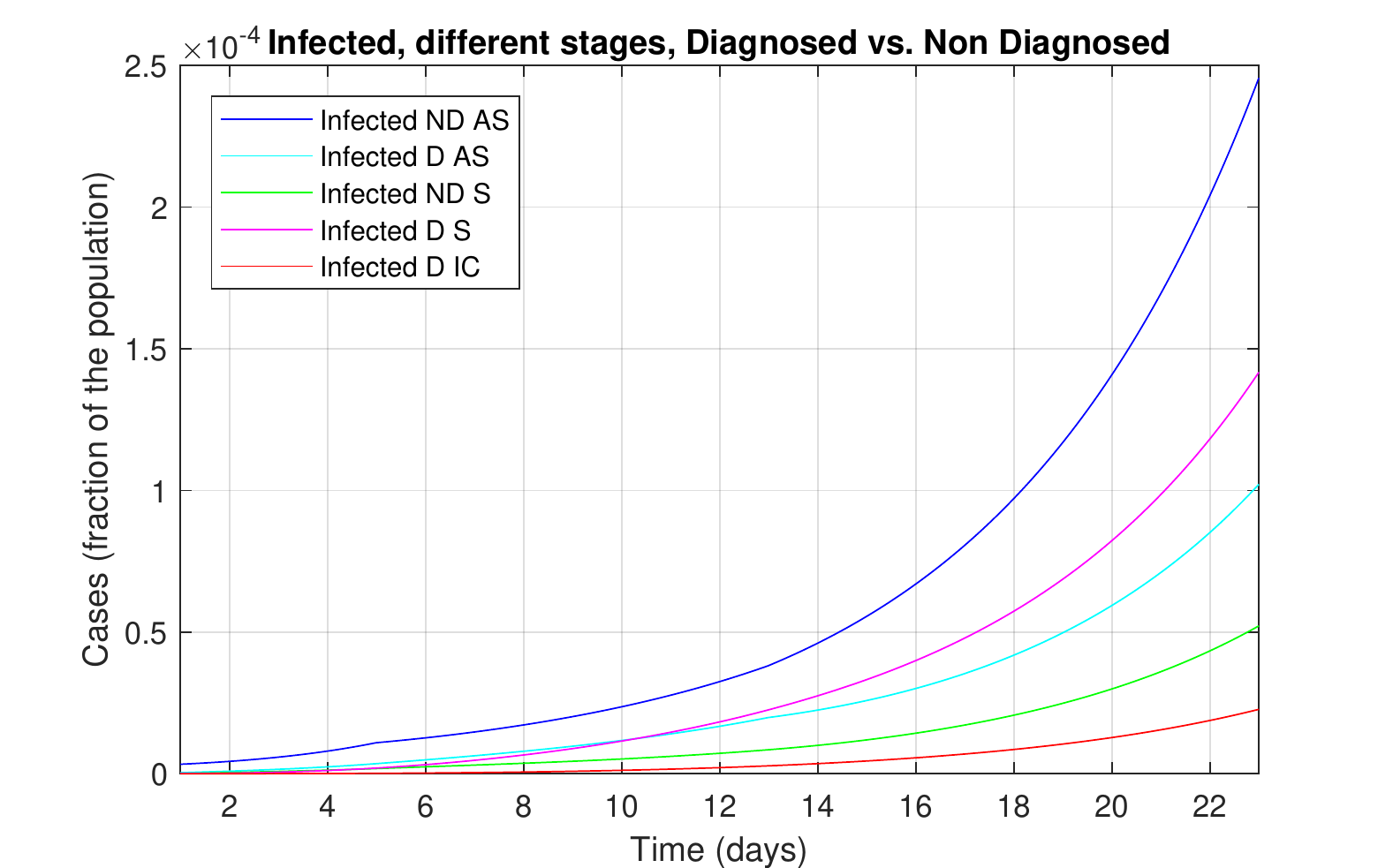}}\\
        \subfloat[Long-term evolution: actual vs. diagnosed cases of infection.]{\includegraphics[width=.48\textwidth]{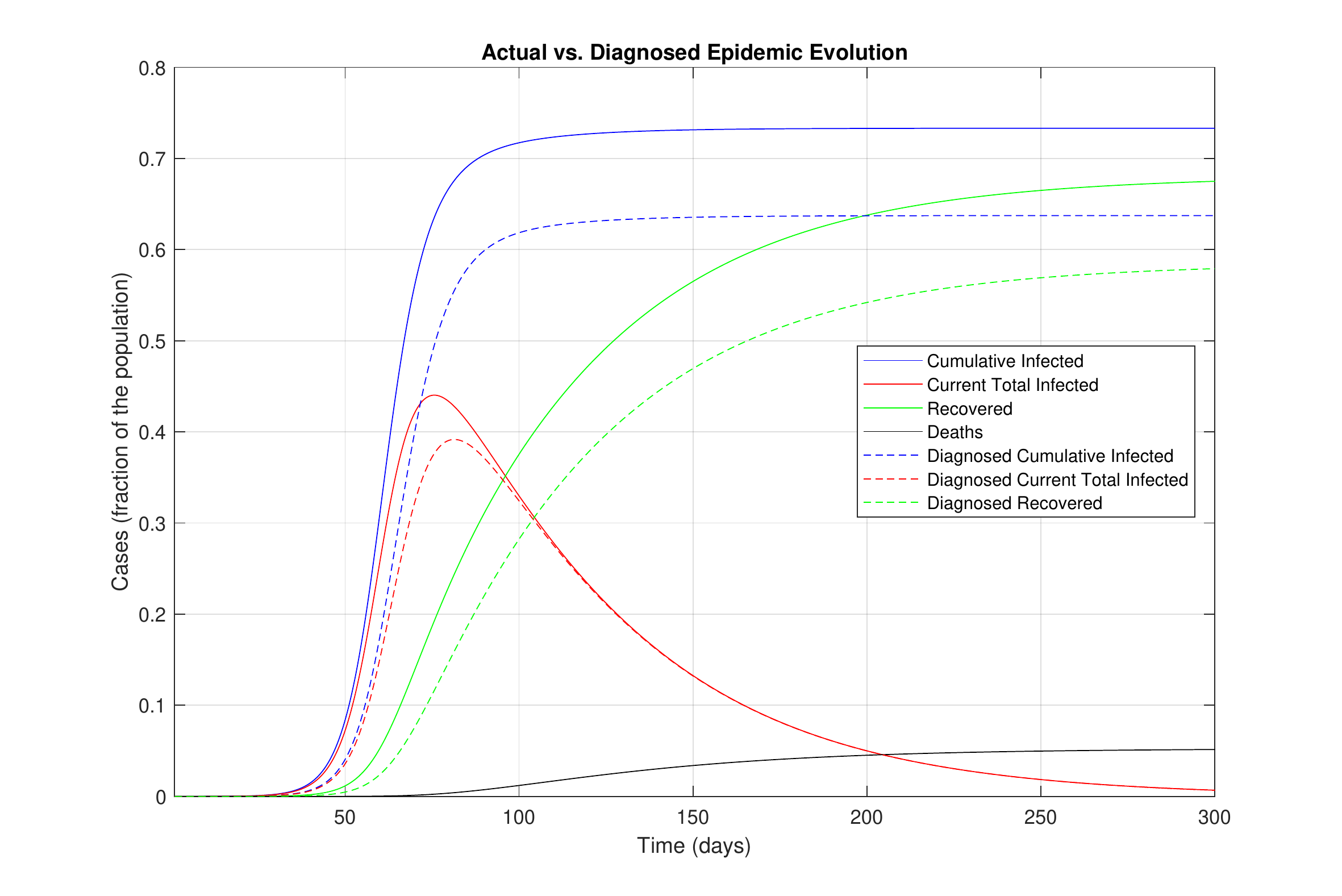}} \quad
        \subfloat[Long-term evolution of infected sub-populations.]{\includegraphics[width=.48\textwidth]{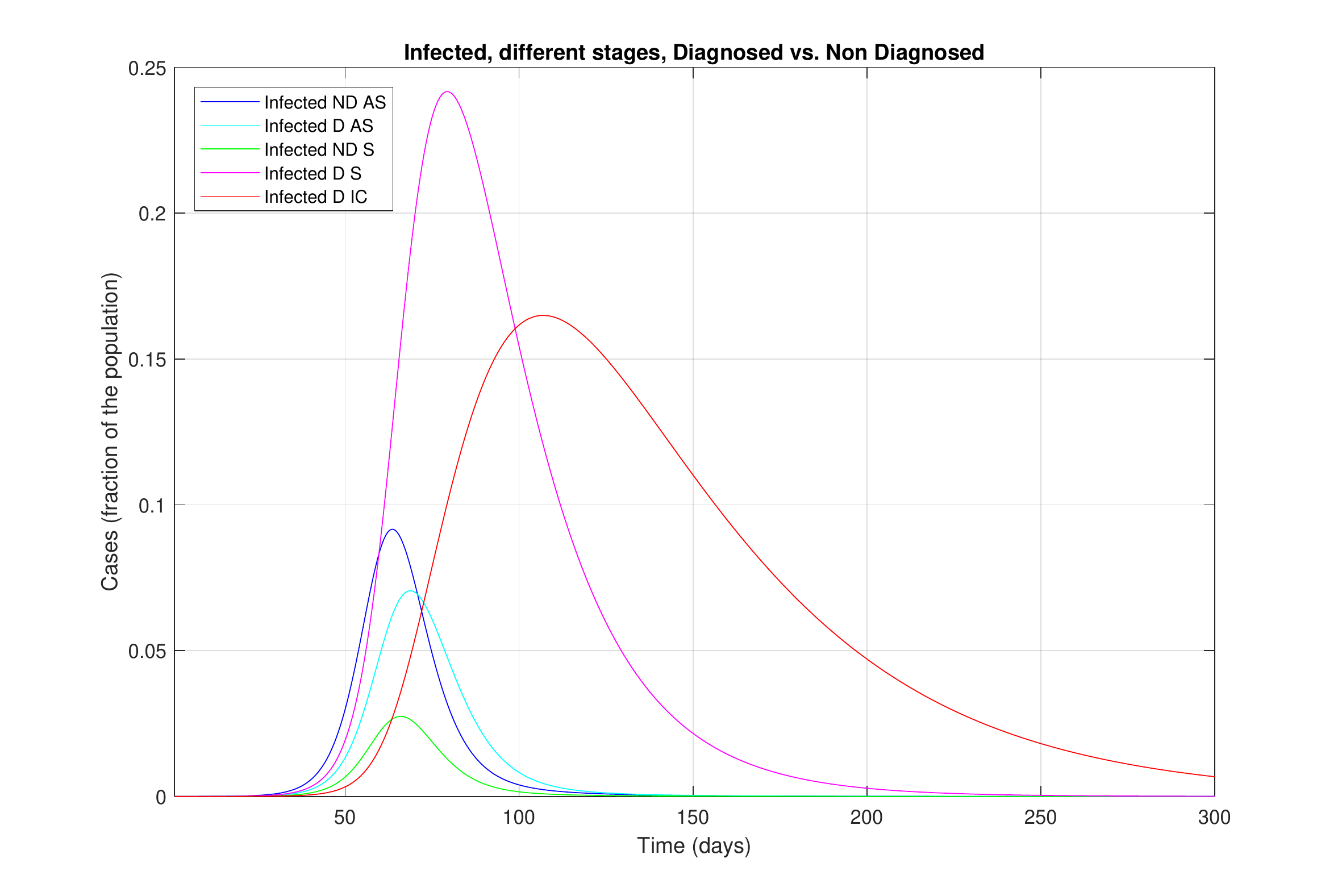}}
                \caption{Epidemic evolution predicted by the model in the absence of further social-distancing countermeasures. The difference between the actual and the perceived evolution of the epidemics is shown; we assume that all deaths caused by the epidemics are correctly identified as such, while the number of infected and recovered individuals is underestimated because not all the population is tested and diagnosed.}
        \label{fig:Case1}
\end{figure}

In the absence of further countermeasures, the model predicts an evolution that leads to $73\%$ of the population having contracted the virus (and about $64\%$ having been diagnosed), and about $5.2\%$ of the population having died because of the contagion over a $300$-days horizon, as shown in Figure \ref{fig:Case1} (c).
Note that the number of eventually recovered people can be computed based on the formula \eqref{R0H}, while the overall number of deaths can be computed as in \eqref{R0E}.
The peak of the number of concurrently infected individuals occurs around $76$ days, and amounts to around $44\%$ of the population; however, the peak of concurrently \emph{diagnosed} infected individuals occurs later, around $82$ days, and amounts to $39\%$ of the population.
The actual CFR is then around $7.2\%$, while the perceived CFR is around $9.0\%$; these values can be computed directly using the formulas in \eqref{eq.barM} and \eqref{eq.barP} respectively. Figure \ref{fig:Case1} (d) shows how the different sub-populations of infected individuals evolve over time, and it is interesting to notice that each sub-population reaches its peak at a different time. In particular, the fraction of infected who need Intensive Care reaches its peak, almost $16.5\%$ of the population, after $107$ days.

Clearly some containment measures are needed, and indeed the Italian government decided for a country-wide lockdown starting with March 9th, whose effect will clearly be visible with a delay. It is still unclear how strongly this lockdown will reduce the infection parameters. 
We consider possible different scenarios.

\begin{figure}[p]
        \centering
        \subfloat[Epidemic evolution: actual vs. diagnosed cases of infection.]{\includegraphics[width=.48\textwidth]{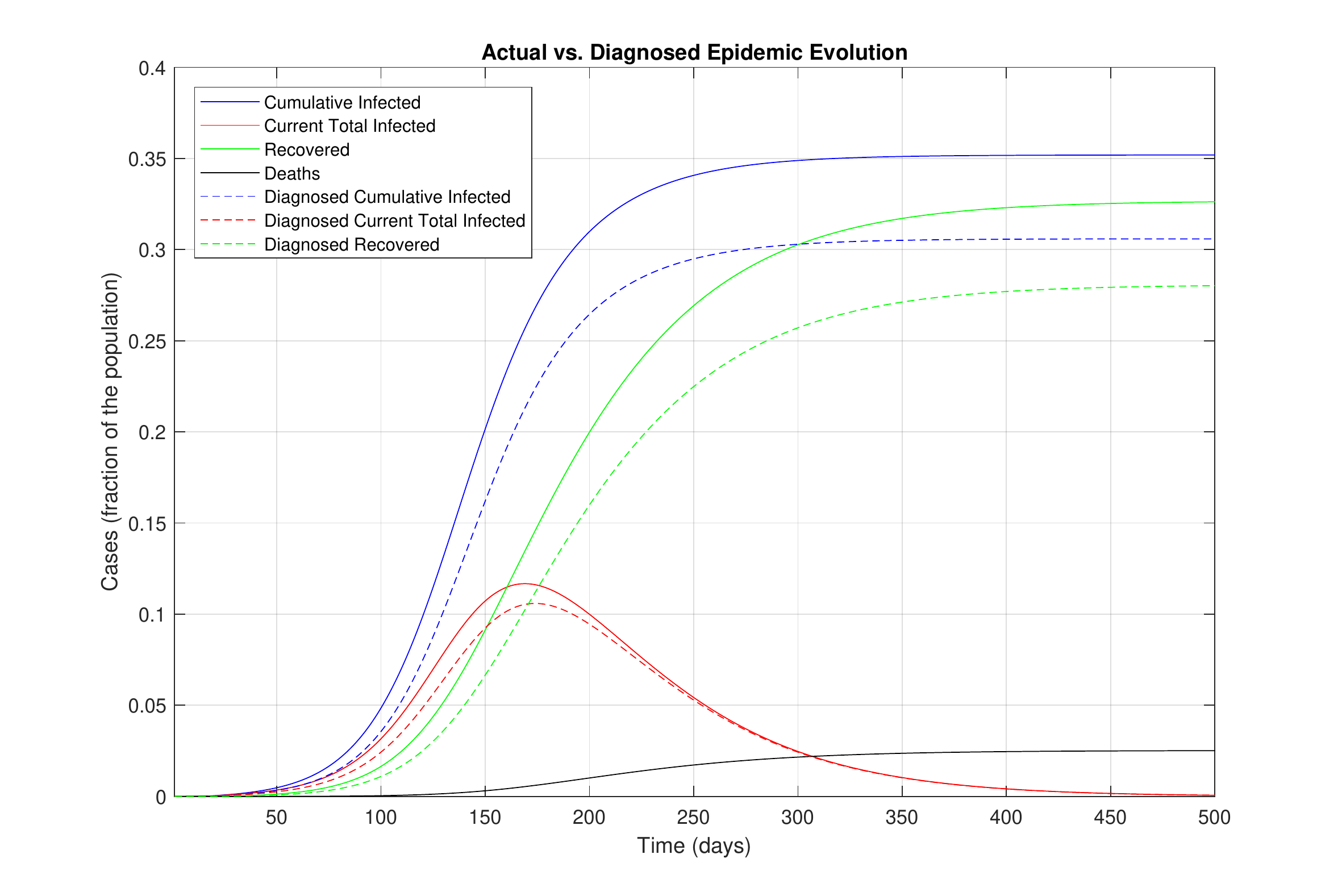}} \quad
        \subfloat[Evolution of infected sub-populations.]{\includegraphics[width=.48\textwidth]{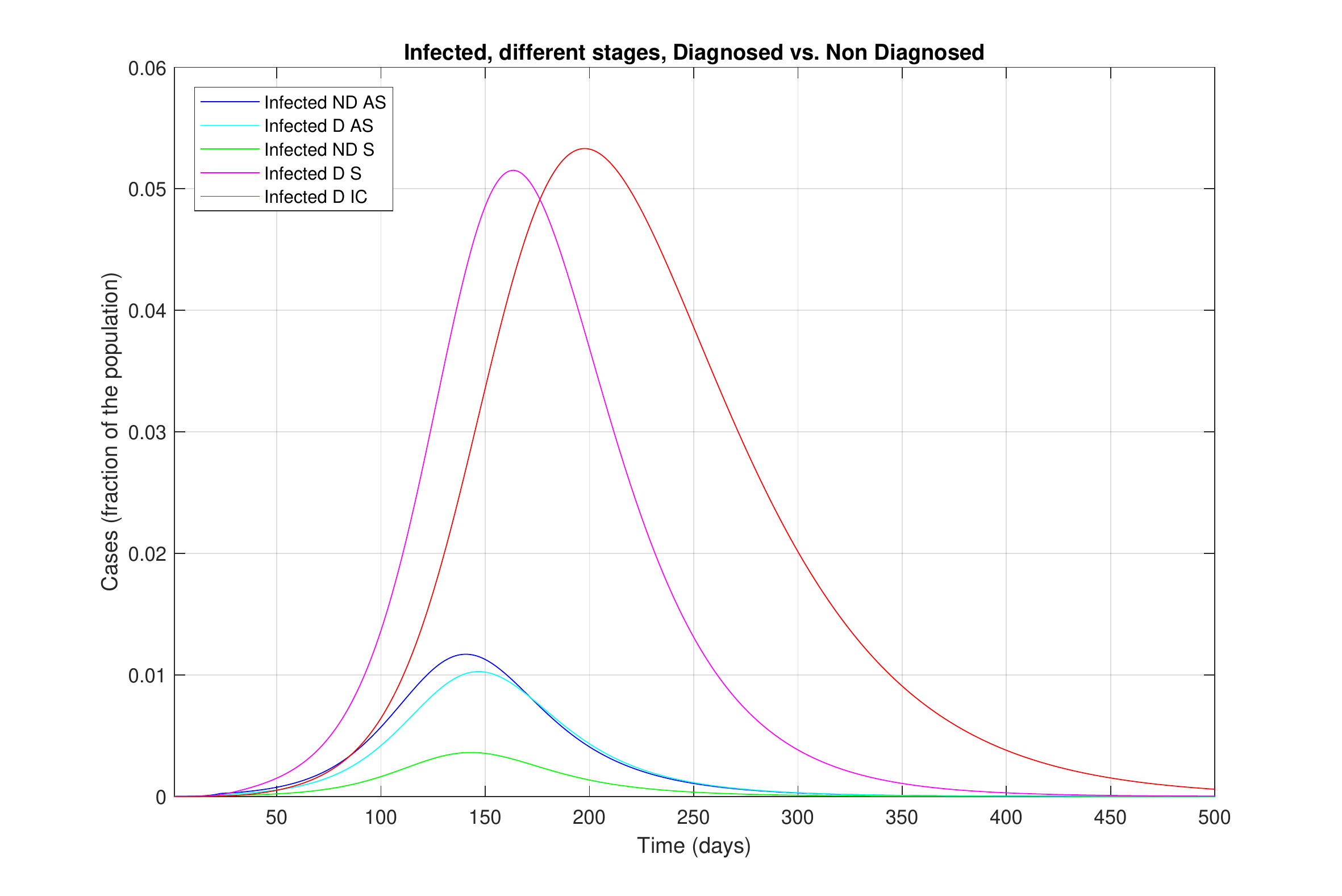}}\\
        \subfloat[Epidemic evolution: actual vs. diagnosed cases of infection.]{\includegraphics[width=.48\textwidth]{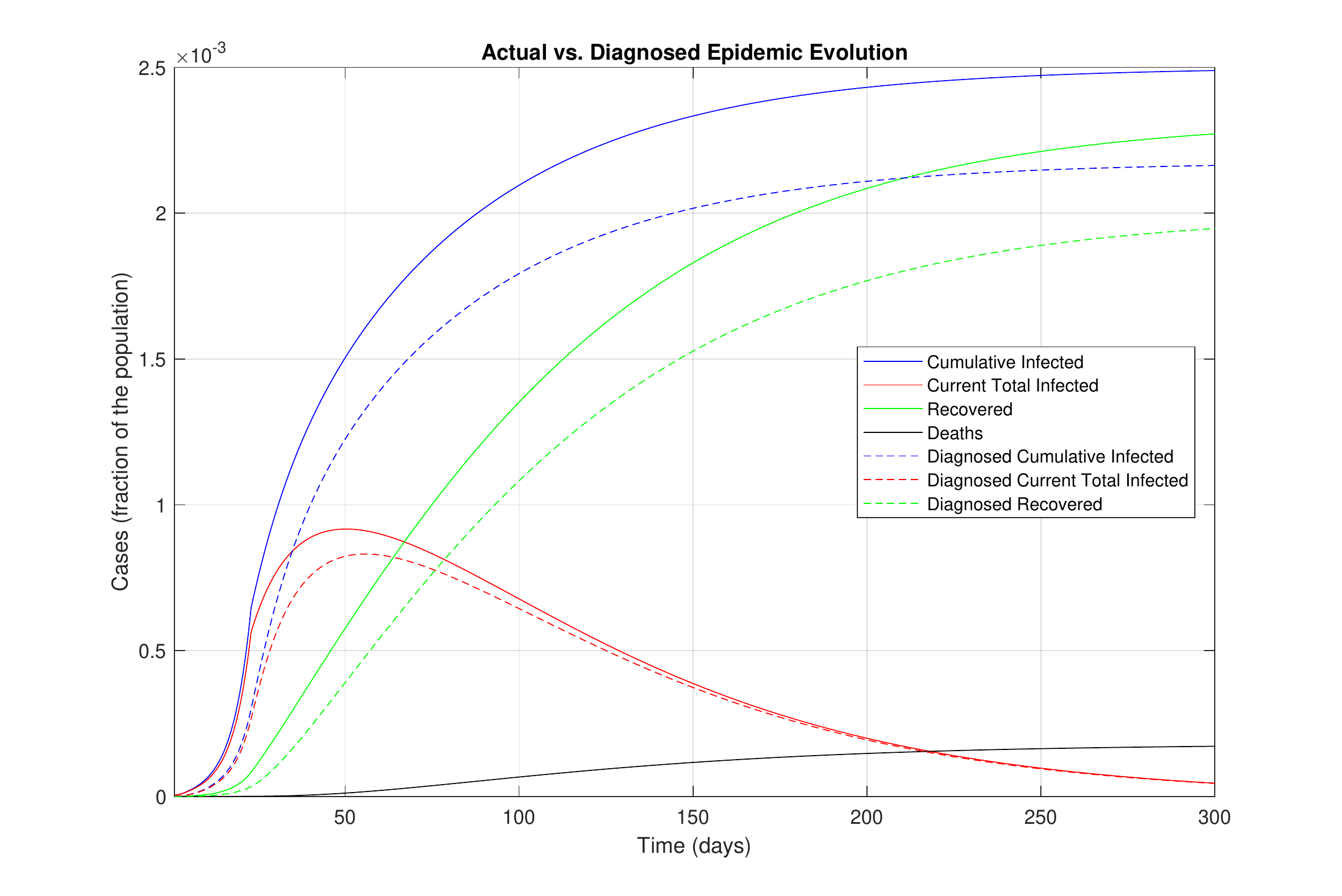}} \quad
        \subfloat[Evolution of infected sub-populations.]{\includegraphics[width=.48\textwidth]{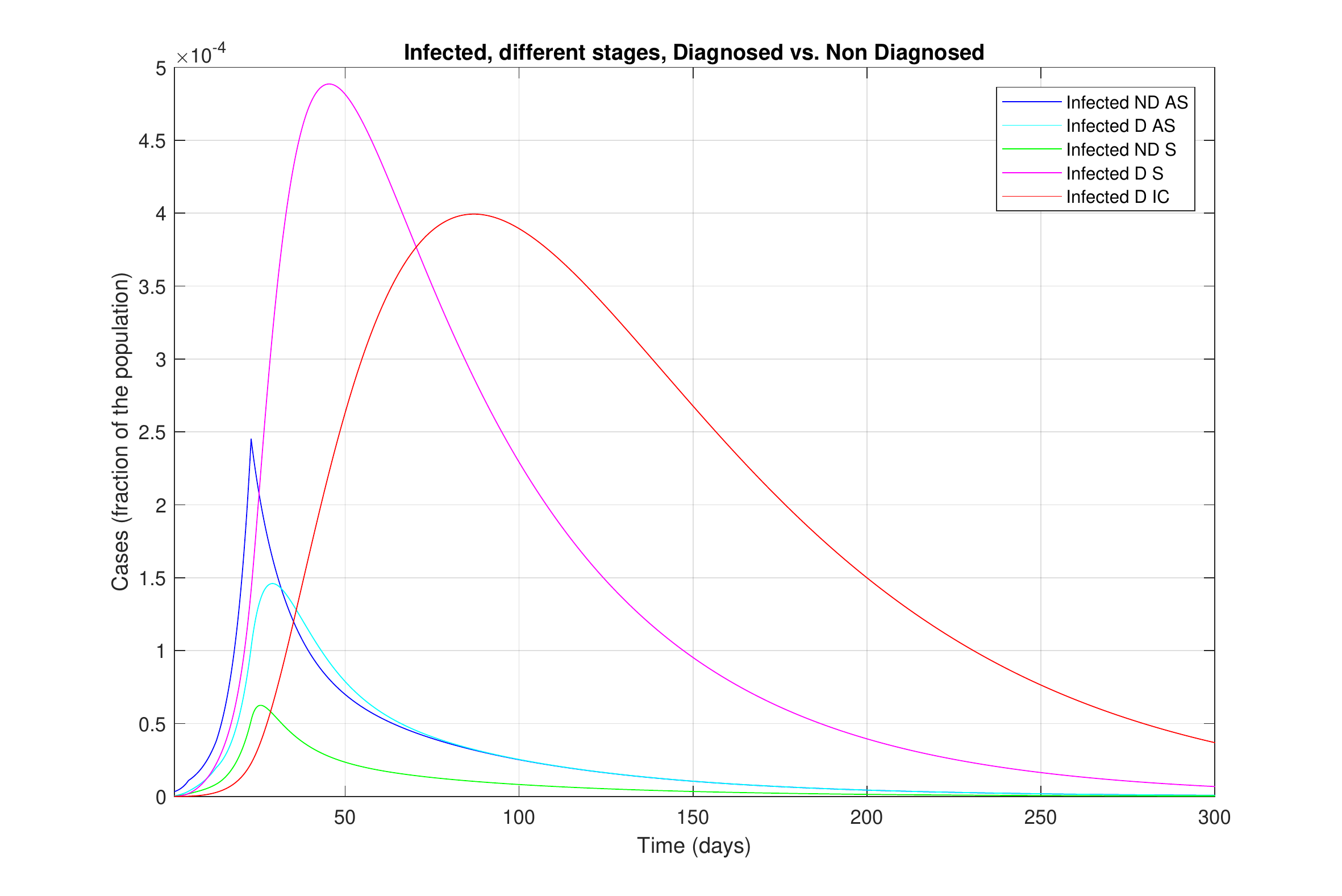}}\\
                \subfloat[Epidemic evolution: actual vs. diagnosed cases of infection.]{\includegraphics[width=.48\textwidth]{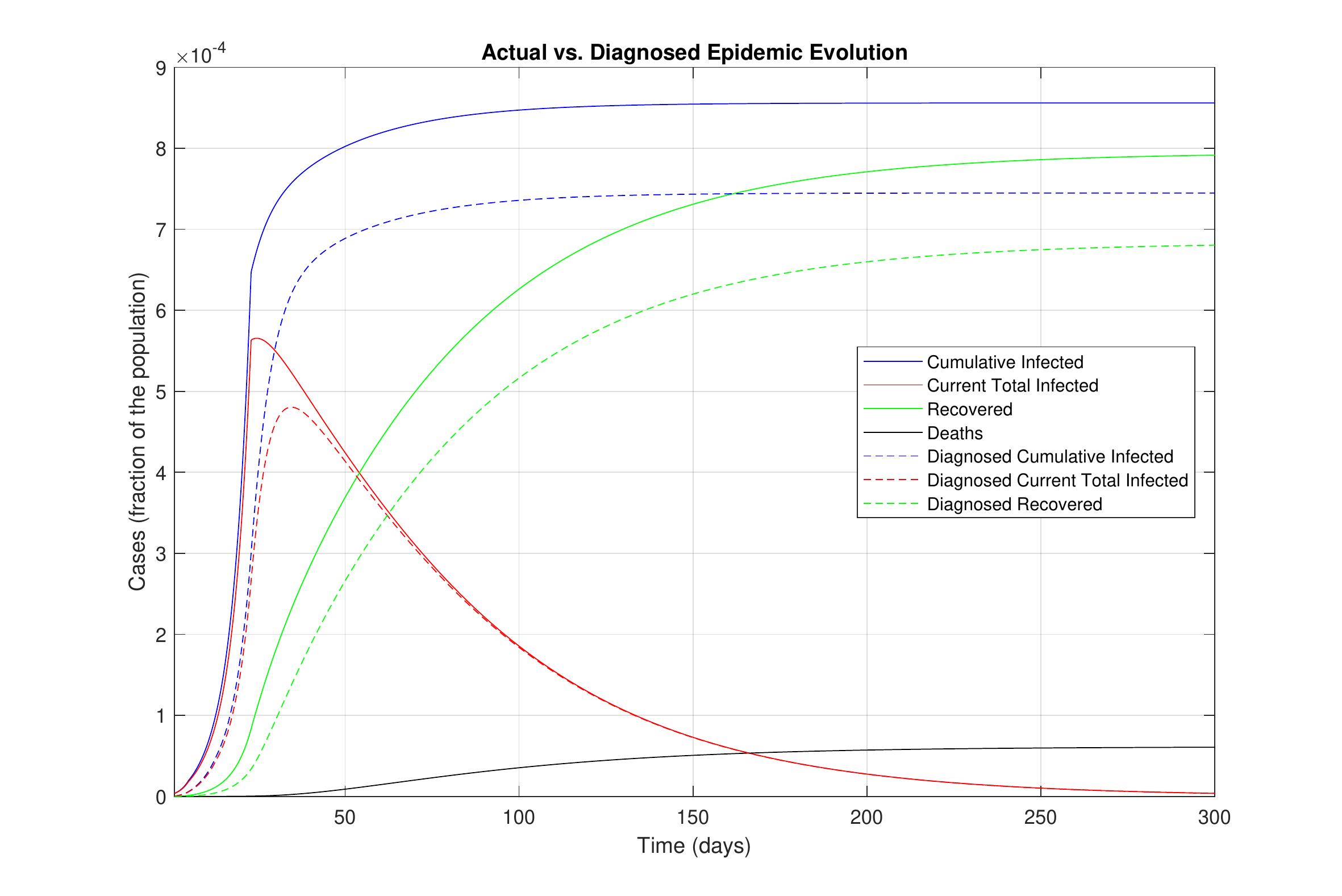}} \quad
        \subfloat[Evolution of infected sub-populations.]{\includegraphics[width=.48\textwidth]{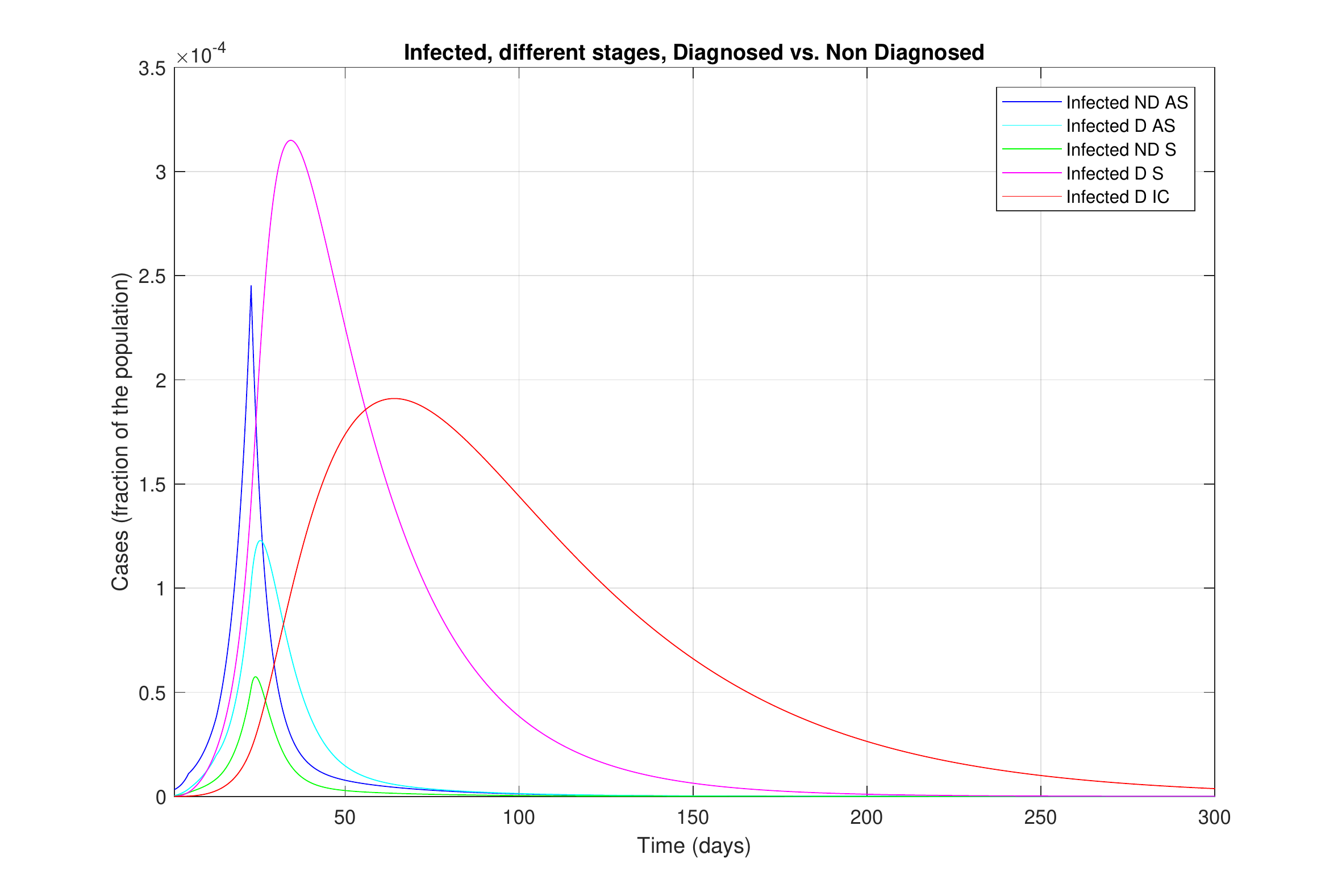}}
                \caption{Epidemic evolution predicted by the model with mild (panels a and b), strong (panels c and d) and very strong (panels e and f) social-distancing countermeasures. The difference between the actual and the perceived evolution of the epidemics is shown; we assume that all deaths caused by the epidemics are correctly identified as such, while the number of infected and recovered individuals is underestimated because not all the population is tested and diagnosed.}
        \label{fig:Case2}
\end{figure}

First, we assume that the increased social-distancing countermeasures lead after day $22$ to $\alpha=0.285$ and $\gamma=0.171$, thus $R_0=1.13$, which is still larger than $1$.
Hence, the peak is delayed (and reduced in amplitude), because the increase in the number of new infected is reduced. This helps gain more time to strengthen and supply the healthcare system, but is not enough to make sure the epidemic phenomenon recedes.
Over a $500$-days horizon, as shown in Figure \ref{fig:Case2} (a), the model predicts an evolution that leads to a peak in the number of concurrently infected individuals around day $170$, amounting to $11.7\%$ of the population ($10.6\%$ of the population have been diagnosed).
Eventually, $35\%$ of the population has contracted the virus (and about $30\%$ have been diagnosed), and about $2.5\%$ of the population has died because of the contagion.
The actual (resp. perceived) CFR is then still around $7.2\%$ (resp. $9.0\%$).
The fraction of patients in need of Intensive Care, as shown in Figure \ref{fig:Case2} (b), reaches its peak on day $198$, amounting to $5.3\%$ of the population.
We see that the adopted social-distancing policy, although mild, have some impact. They are still insufficient, though.

Stronger social-distancing countermeasures are required, able to yield, after day $22$, $\alpha=0.200$ and $\gamma=0.086$, hence $R_0=0.787$, which is now lower than $1$.
Now the peak is not delayed, but anticipated, because the increase in the number of new infected is reduced so much that it soon becomes a decrease.
Over a $300$-days horizon, as shown in Figure \ref{fig:Case2} (c), the model predicts an evolution of the situation that leads to a peak in the number of concurrently infected individuals around day $50$, amounting to $0.092\%$ of the population; the peak in \emph{diagnosed} infected occurs at day $54$ and amounts to $0.083\%$ of the population.
Eventually, $0.25\%$ of the population has contracted the virus (and about $0.22\%$ have been diagnosed), and about $0.02\%$ of the population has died because of the contagion.
The actual (resp. perceived) CFR is then still around $7.2\%$ (resp. $9.0\%$).
The fraction of patients in need of Intensive Care, as shown in Figure \ref{fig:Case2} (d), reaches its peak on day $85$, amounting to $0.04\%$ of the population.

Even stronger social-distancing countermeasures are worth being adopted: after day $22$, $\alpha=\gamma=0.057$, hence $R_0=0.329$, which is significantly lower than $1$.
Over a $300$-days horizon, as shown in Figure \ref{fig:Case2} (e), the model predicts an evolution of the situation that leads to a peak in the number of concurrently infected individuals around day $25$, amounting to $0.057\%$ of the population; the peak in \emph{diagnosed} infected occurs at day $35$ and amounts to $0.048\%$ of the population.
Eventually, $0.086\%$ of the population has contracted the virus (and about $0.074\%$ have been diagnosed), and about $0.006\%$ of the population has died because of the contagion.
The actual (resp. perceived) CFR is then still around $7.2\%$ (resp. $9.0\%$).
The fraction of patients in need of Intensive Care, as shown in Figure \ref{fig:Case2} (f), reaches its peak on day $64$, amounting to $0.02\%$ of the population.

\begin{figure}[htb]
        \centering
        \subfloat[Currently infected: $D(t)+R(t)+T(t)$.]{\includegraphics[width=.48\textwidth]{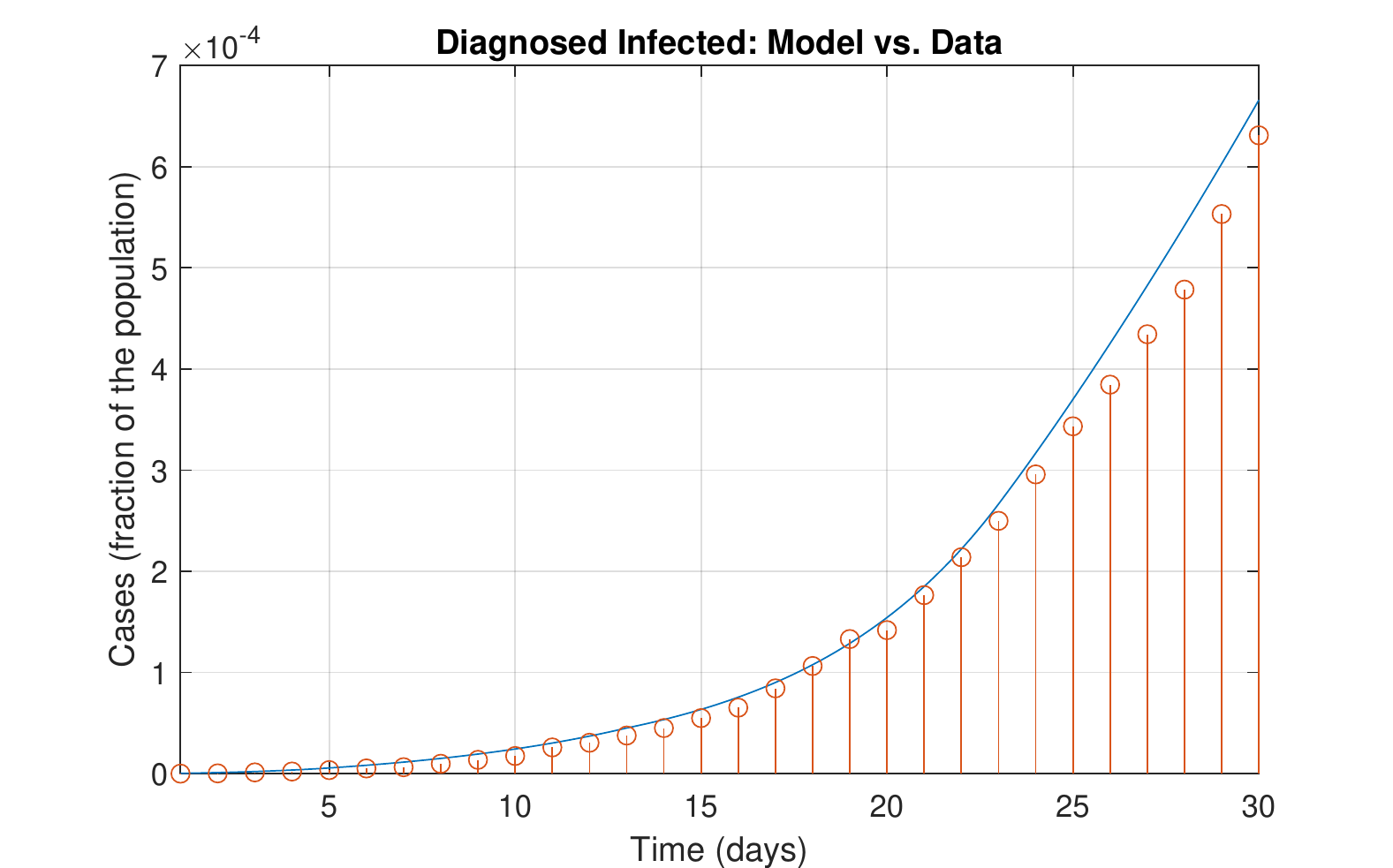}}
        \subfloat[Cumulative diagnosed cases: {$D(t)+R(t)+T(t)+E(t)+\int_0^t (\rho D(\phi)+ \xi R(\phi)+\sigma T(\phi))d\phi$.}]{\includegraphics[width=.48\textwidth]{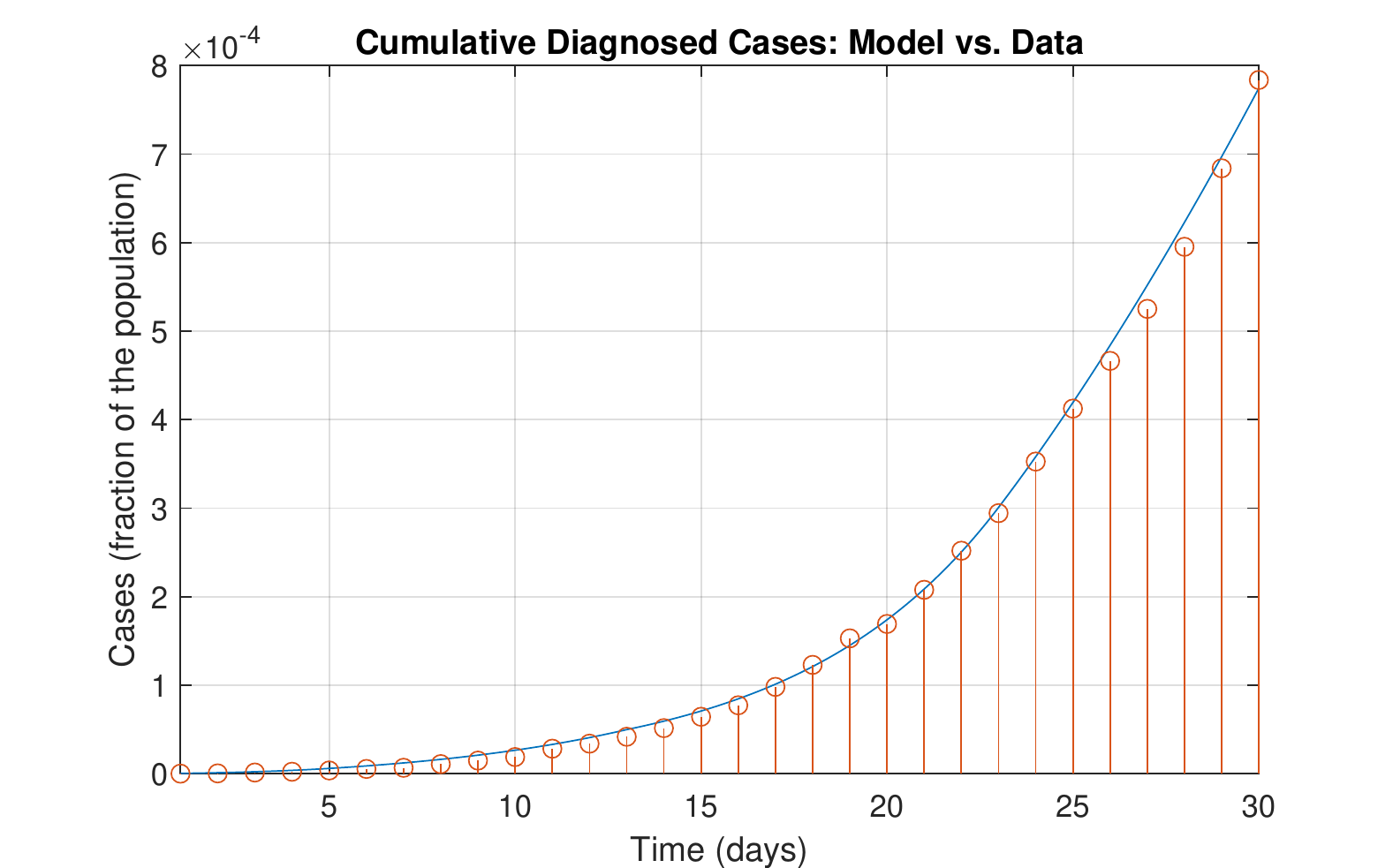}} \quad
                \caption{Comparison between the official data (red dots histogram) and the results with the calibrated SIDARTHE model in the presence of mild social-distancing countermeasures (blue line).}
        \label{fig:forecast}
\end{figure}

\paragraph{Forecast.} We have fitted the model based on the available data in the period February 20, 2020 (day $1$) to March 12, 2020 (day $22$). Now the data up to March 20, 2020 (day $30$) are available. Which is the scenario they fit the best? Concerning  the number of cumulative diagnosed cases and the number of currently infected, the best fit is with the scenario with mild social-distancing countermeasures (discussed in Figure \ref{fig:Case2}, panels a and b), as can be seen in Figure \ref{fig:forecast}. Of course the effect of the adopted social-distancing policies can be seen with some delay, so we can hope that the scenario will get closer to the one with strong social-distancing countermeasures in the near future.
If not, our prediction suggests the need of even stronger countermeasures.

\section{Discussion}

The key feature of our proposed model is the distinction between detected and undetected infection cases, and between cases with different SOI classification (mild and moderate vs. major and extreme).

Distinguishing between diagnosed and not diagnosed cases allows us to highlight the perceived distortion in disease statistics, such as the number of infected individuals, the transmission rate and the case fatality rate (CFR, typically defined as the ratio between the number of deaths ascribed to the infection and the number of diagnosed cases). The discrepancy between the actual CFR (total number of deaths due to the infection, divided by the total number of people who have been infected) and the perceived CFR (number of deaths ascribed to the infection, divided by the number of people who have been diagnosed as infected) can be quantified based on this model.
Therefore, the model can explain the possible discrepancy between the actual infection dynamics and the perception of the phenomenon. Misperception (either resulting in underestimating or overestimating) can be particularly relevant in the early phases of an epidemic phenomenon, due to the lack of thorough information: for instance, performing an insufficient number of tests may lead to underestimating the transmission rate (because many infected subjects are not diagnosed as such) and overestimating the case fatality rate (because critical or fatal cases hardly go undetected).
The model thus provides a rough quantification of the error in estimating the actual number of infected people due to the lack of proper diagnostic tests, or due to insufficient number of diagnostic tests being performed. Also, it can explain and predict the long-term effects of underdiagnosis, including the (apparently surprising) increased number of infections and fatalities, with sudden outbreaks after long silent periods.

Concerning diagnostic tests for COVID-19, currently, standard molecular methods to detect the presence of SARS-CoV-2 in respiratory samples are aspecific real-time reverse transcriptase polymerase chain reaction (RT-PCR), which target RNA dependent RNA polymerase and E genes \cite{Corman2020}.
These are time consuming techniques that cannot be performed to all the susceptible population. Moreover, high false negatives rates are reported and certified laboratories with expensive equipment are needed \cite{Gallagher}.
Rapid tests with high sensitivity and specificity are urgently required to be easily developed in the real life settings such as schools, airports, train stations etc. Some laboratories are actually moving in this direction by developing a 15 minutes test to detect SARS-CoV-2 IgM and IgG simultaneously in human blood \cite{Li2020}.

Once the model parameters have been estimated on the basis of the available clinical data, the model enables us to reproduce and predict the dynamic evolution of the epidemic, and to evaluate the possible underestimation or overestimation of the epidemic phenomenon based on current statistics, which are heavily subject to bias (e.g, asymptomatic patients may get tested according to some protocols, not tested according to others).
The model helps evaluate and predict the effect of the implementation of different guidelines and protocols (for example, more extensive screening for the disease or stricter social distancing measures), which typically results in a change in the model parameters.
In general, we can observe the following.
\begin{itemize}
\item The model predictions in the long run are not very sensitive to the initial conditions, but they are extremely sensitive to the parameter values, which are deeply uncertain and can vary due to several factors, such as population density, cultural habits, environmental conditions, age distribution of the population.
\item The predictions must take into account parameter variations due to the measures imposed by the Government. This is a fundamental aspect: in the long term, not imposing drastic measures leads to catastrophic outcomes, even when the initially affected population is a small fraction.
\item  Social-distancing measures are modelled by reducing the infection coefficients ($\alpha$, $\beta$, $\gamma$ and $\delta$). The infection peak time is not monotonic
with increasing restrictions. Partial restrictions on the population movements initially postpone the peak, while strong restrictions anticipate the peak. Mild containment measures may have negative effects, for instance augmenting the fraction of population with life-threatening symptoms with respect to the fraction of population with mild symptoms.
\item Diagnosis campaigns can reduce the infection peak, because the diagnosed population enters the quarantine, hence it is less likely to affect the susceptible population. Anyway, healthcare workers are particularly exposed and their risk of infection is ever increasing, as supported by Chinese reports for the COVID-19 case \cite{Roosa2020,CWang2020} that suggest disease amplification in healthcare settings despite restrictive measures.
\item The model does not take into account reduced availability of medical care due to the healthcare system reaching or even surpassing its capacity \cite{Ji2020}. This can be done only indirectly. For instance, when the number of seriously affected individuals is high (above a threshold), the mortality coefficient can be increased, due to the lack of ICUs.
\end{itemize}

Our simulation results, achieved by combining the model with the available data about the COVID-19 epidemic in Italy, suggest the urgent need -- and the effectiveness -- of strong social-distancing measures, in line with \cite{Wu2020,Casella2020,Hellewell2020}.

We compare scenarios with control measures of varying strength, predicting for each the timing and magnitude of the epidemic peak, including the peak of ICU admissions. According to our findings, a partial implementation of control measures will result in a delay in the peak of infected individuals and patients admitted to the ICU, contrasting with an only moderate decrease in the total number of infected individuals and ICU admission due to COVID-19. 
On the other hand, the implementation of very strong social-distancing strategies will result in an anticipated peak of infected individuals and patients admitted to the ICU, with a marked decrease in the total number of infected individuals and ICU admissions due to the disease, and is by and large the preferable option.

Although the mortality rate (number of deaths in the whole population) of COVID-19 is a decreasing function of restriction measures, the case fatality rate (number of deaths in the infected population) is essentially constant in different scenarios, hence it is unaffected by the extent of social restriction. Despite rigid isolation policies, COVID-19 patients may still be burdened with excess case fatality.
In this perspective, efforts should be placed on developing more effective treatment strategies to combat the disease. As new drugs and vaccines are being tested and evaluated, the current scenario will most likely evolve based on the ongoing innovations in the therapeutic and immunological fields \cite{MWang2020,Chang2020,Diao2020,Chen2020}.

As a possible future development, with the appropriate data, the model can be extended to predict the simultaneous evolution of other diseases, which, due to the epidemic emergency, may be overestimated, underestimated or not treated appropriately because the healthcare system is currently overloaded, thus leading to an increased number of ``collateral'' deaths not directly linked to the virus.

\section{Methods}

\subsection{Equilibrium points}\label{sec:eq}
The possible equilibria are given by $(\bar S,0,0,0,0,\bar T,\bar H)$, with $\bar S+\bar T + \bar  H=1$. Indeed, in the system \eqref{eq:S}-\eqref{eq:E}, $\dot S=0$ if and only if either (i) $S=0$ or (ii) $\alpha I+\beta D+\gamma A+\delta R =0$, namely $I=D=A=R=0$ (or both).

(i) If $S=0$, $\dot I=0$ if and only if $I=0$. Then, $\dot D=0$ if and only if $D=0$ and $\dot A=0$ if and only if $A=0$. Then, $\dot R=0$ if and only if $R=0$. Finally, $\dot T=0$ and $\dot H=0$.

(ii) If $I=D=A=R=0$, then $\dot I=\dot D = \dot A = \dot R = 0$, and also $\dot T=0$ and $\dot H=0$.

\subsection{Proof of Proposition \ref{stab}}\label{sec:stab}
The dynamical matrix of the linearised system around the equilibrium $(\bar S,0,0,0,0,\bar T,\bar H)$ is
\begin{equation}
J = \begin{bmatrix}
0 & -\alpha \bar S & -\beta \bar S & -\gamma \bar S & -\delta \bar S & 0 & 0 & 0\\
0 & \alpha \bar S-r_1 & \beta \bar S & \gamma \bar S & \delta \bar S & 0 & 0 & 0\\
0 & \epsilon & -r_2 & 0 & 0 & 0 & 0& 0\\
0 & \zeta & 0 & -r_3 & 0 & 0 & 0& 0\\
0 & 0 & \eta & \theta & -r_4 & 0 & 0& 0\\
0 & 0 & 0 & \mu & \nu & -r_5 & 0& 0\\
0 & \lambda & \rho & \kappa & \xi & \sigma & 0& 0\\
0 & 0 & 0 & 0 & 0 & \tau &0 &0
\end{bmatrix}
\end{equation}
where $r_1=\epsilon+\zeta+\lambda$, $r_2=\eta+\rho$, $r_3=\theta+\mu+\kappa$, $r_4=\nu+\xi$, $r_5 = \sigma+\tau$. 

The matrix has three null eigenvalues, and four eigenvalues roots of the polynomial
\begin{eqnarray}
\label{eq.pol}
p(s)&=&D(s)-\bar S N(s),
\end{eqnarray}
where 
\begin{align*}
&D(s)= (s+r_1)(s+r_2)(s+r_3)(s+r_4)(s+r_5)\\
&N(s)= (s+r_5)\left\{\alpha (s+r_2)(s+r_3)(s+r_4)+\beta\epsilon(s+r_3)(s+r_4)+\gamma \zeta(s+r_2)(s+r_4)+\delta[\eta\epsilon(s+r_3)+\zeta\theta(s+r_2)]\right\}
\end{align*}

The transfer function from $u$ to $y_S$ in Figure \ref{Fig_intrinsic_feedback} is $G(s)=N(s)/D(s)$. Since the system is positive, the $H_\infty$ norm of $G(s)$ is equal to the static gain $G(0)=N(0)/D(0)$.

Then, by standard root locus (small gain argument) on the positive system $G(s)$, we can say that the polynomial is Hurwitz (all roots in the left hand plane) iff
\begin{eqnarray*}
\bar S<\bar S^*=\frac{r_1r_2r_3r_4}{\alpha r_2r_3r_4+\beta\epsilon r_3r_4+\gamma\zeta r_2r_4+\delta(\eta\epsilon r_3+\zeta\theta r_2)},
\end{eqnarray*}
where $\bar S^* = 1/G(0)$.
Therefore we are well justified to define the basic reproduction parameter 
\begin{equation*}
R_0:=\frac{1}{\bar S^*}=
\frac{\alpha +\beta\epsilon/r_2+\gamma\zeta/r_3+\delta(\eta\epsilon/ (r_2r_4)+\zeta\theta/(r_3r_4))}{r_1}
\end{equation*} 
and stability of the equilibrium occurs for $\bar S\bar R_0<1$.

Notice that $R_0=G(0)$ is the $H_\infty$ norm of the transfer function $G(s)$.

\subsection{Proof of Proposition \ref{pro:threshold}}\label{sec:thresh}
Since $S(t)$ is monotonically decreasing and non-negative, it has a limit $\bar S \geq 0$. For $t$ large enough, we have $S(t) \approx \bar S$.
Then the feedback system converges to the system with the static feedback $\bar S$. If, by contradiction, $\bar S$ renders this system unstable, then $x(t)$ diverges, since the Metzler matrix $F+b\bar S c^\top$ has a positive dominant eigenvalue. In turn, this implies that $x(t)$ cannot converge to zero, hence its components remain positive, which means that $\alpha I + \beta D + \gamma A + \delta R >0$ does not converge to zero. As a consequence, also $\dot S = -S(\alpha I + \beta D + \gamma A + \delta R) < 0$ does not converge to zero, hence $S(t)$ cannot converge to a non-negative value $\bar S \geq 0$. We have reached a contradiction.

\subsection{Proof of Proposition \ref{pro:steady}}\label{sec:formula}
From \eqref{eq:S_y}, we have $\dot S(t)/S(t)=-y_S(t)$, namely $-y_S(t)=\frac{d\log(S(t))}{dt}$. By integration we have
$$
\int_0^\infty y_S(\phi) d\phi = -\log \left ( \frac{\bar S}{S(0)}\right ) = \log \left ( \frac{S(0)}{\bar S}\right )
$$
Now, assuming that $F$ and $b$ are constant, we integrate $\dot x(t)$:
$$
 \int_0^\infty \dot x(\phi)d\phi = x(\infty)-x(0) = F \int_0^\infty x(\phi) d\phi + b \int_0^\infty u(\phi) d\phi =  F \int_0^\infty x(\phi) d\phi + b\int_0^\infty S(\phi) y_S(\phi) d\phi.
$$
Since $\dot S(t) = -S(t)y_S(t)$ and $x(\infty) =0$, we have
$$
-x(0) = F \int_0^\infty x(\phi) d\phi - b\int_0^\infty \dot S(\phi) d\phi = F \int_0^\infty x(\phi) d\phi - b(\bar S -S(0)).
$$
We pre-multiply by $c^\top F^{-1}$ and take into account that $y_S(t)=c^\top x(t)$:
$$
-c^\top F^{-1}x(0) = \int_0^\infty y_S(\phi) d\phi - c^\top F^{-1}b (\bar S -S(0)) = \log \left ( \frac{S(0)}{\bar S}\right ) - c^\top F^{-1}b (\bar S -S(0)).
$$
A simple calculation shows that $-c^\top F^{-1}b=R_0$, with $R_0$ defined as in \eqref{eq.R0}. Denoting $f_0=-c^\top F^{-1}x(0)$, we have
$$
f_0  + R_0 (S(0)-\bar S ) =\log \left ( \frac{S(0)}{\bar S}\right )
$$
The other two formulas for $\bar H$ and $\bar E$ in Proposition 3 can be obtained by pre-multiplying the expression of $\int_0^\infty x(\phi) d\phi $ above by $f^\top$ and $d^\top$, respectively.

\subsection{Fit of the Model for the COVID-19 Outbreak in Italy}

\begin{figure}[p]
        \centering
        \subfloat[Currently infected without symptoms: $D(t)$.]{\includegraphics[width=.49\textwidth]{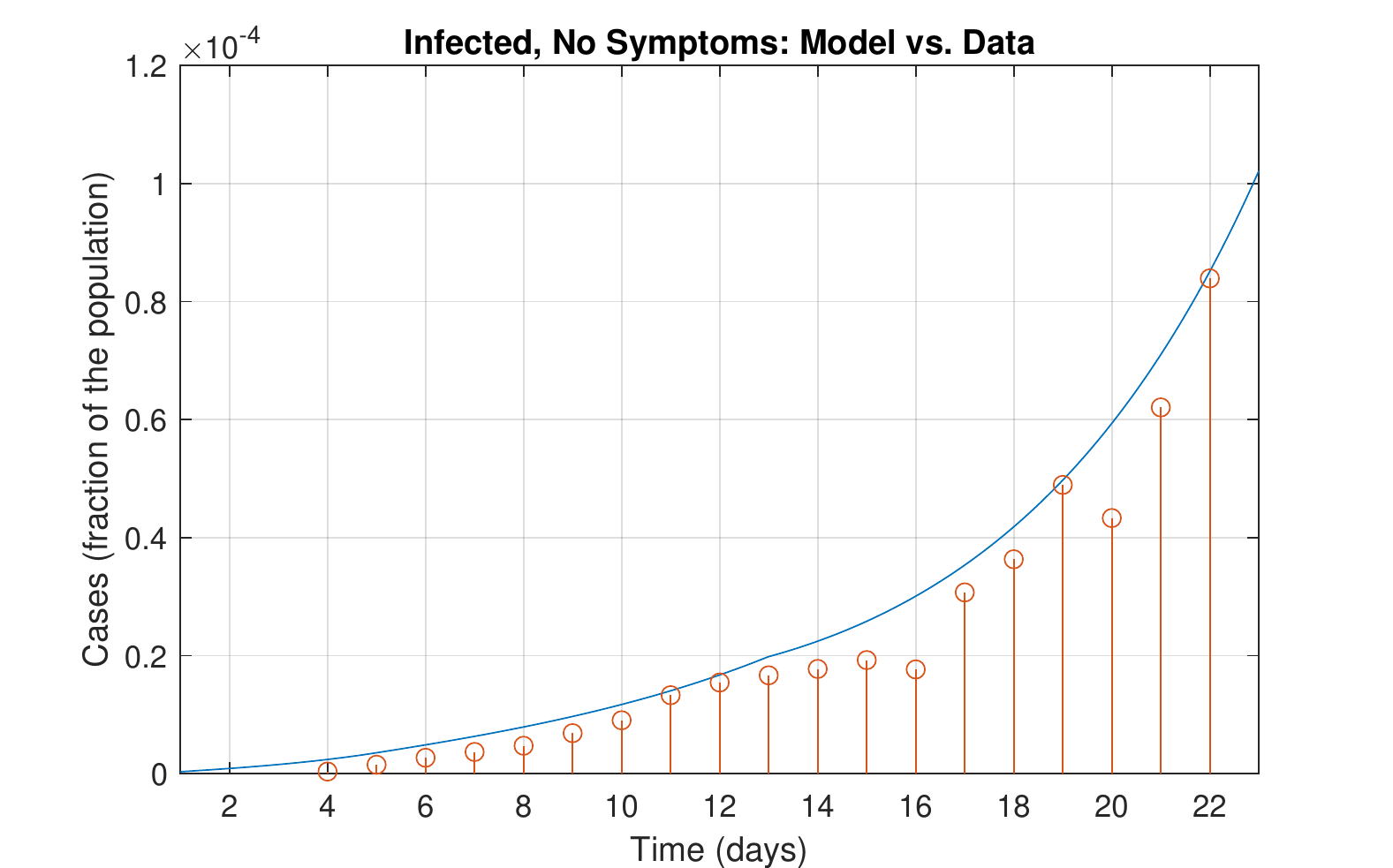}}
                \subfloat[Currently infected with symptoms: $R(t)$.]{\includegraphics[width=.49\textwidth]{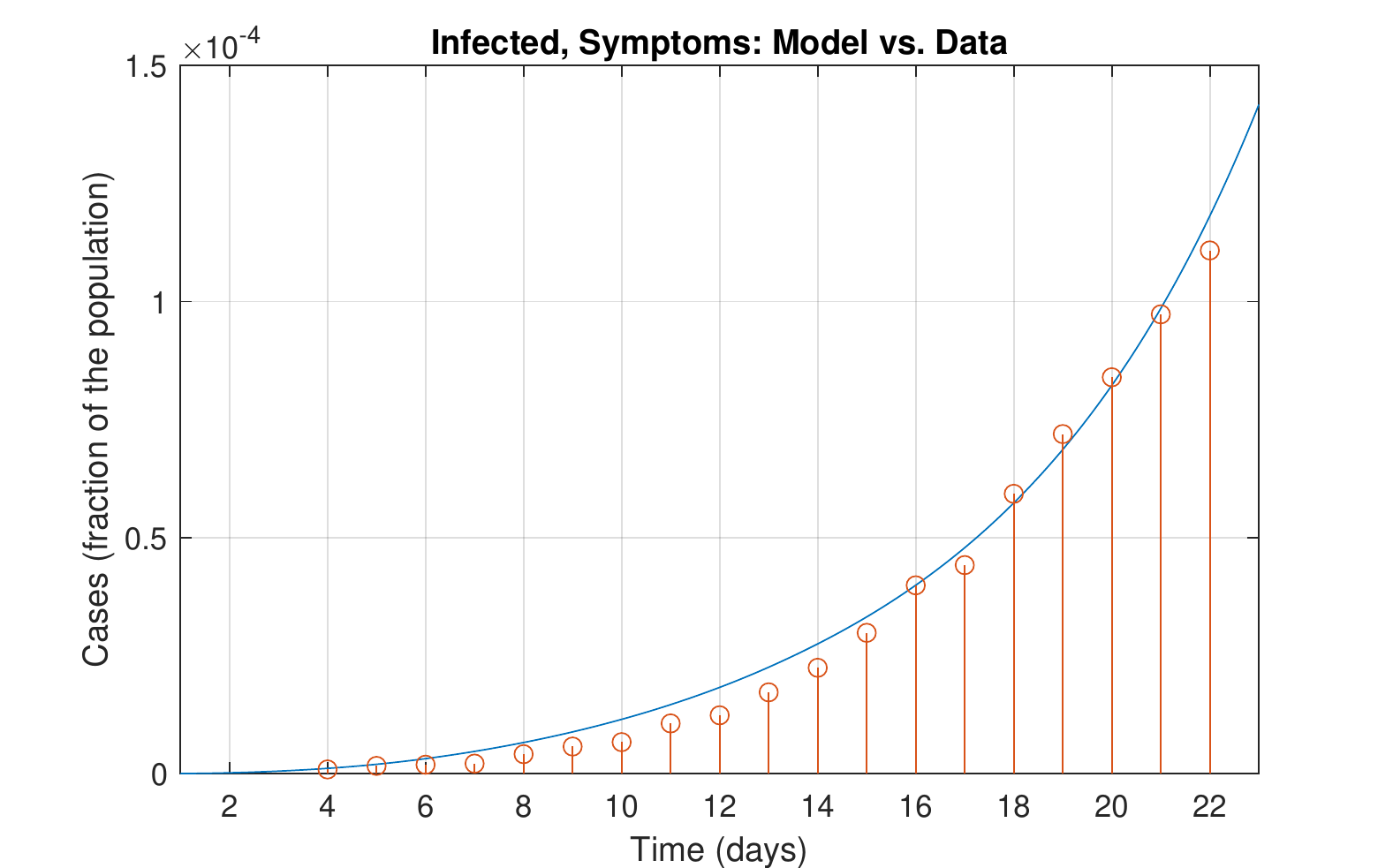}}\\
                \subfloat[Currently infected with life-threatening symptoms: $T(t)$.]{\includegraphics[width=.49\textwidth]{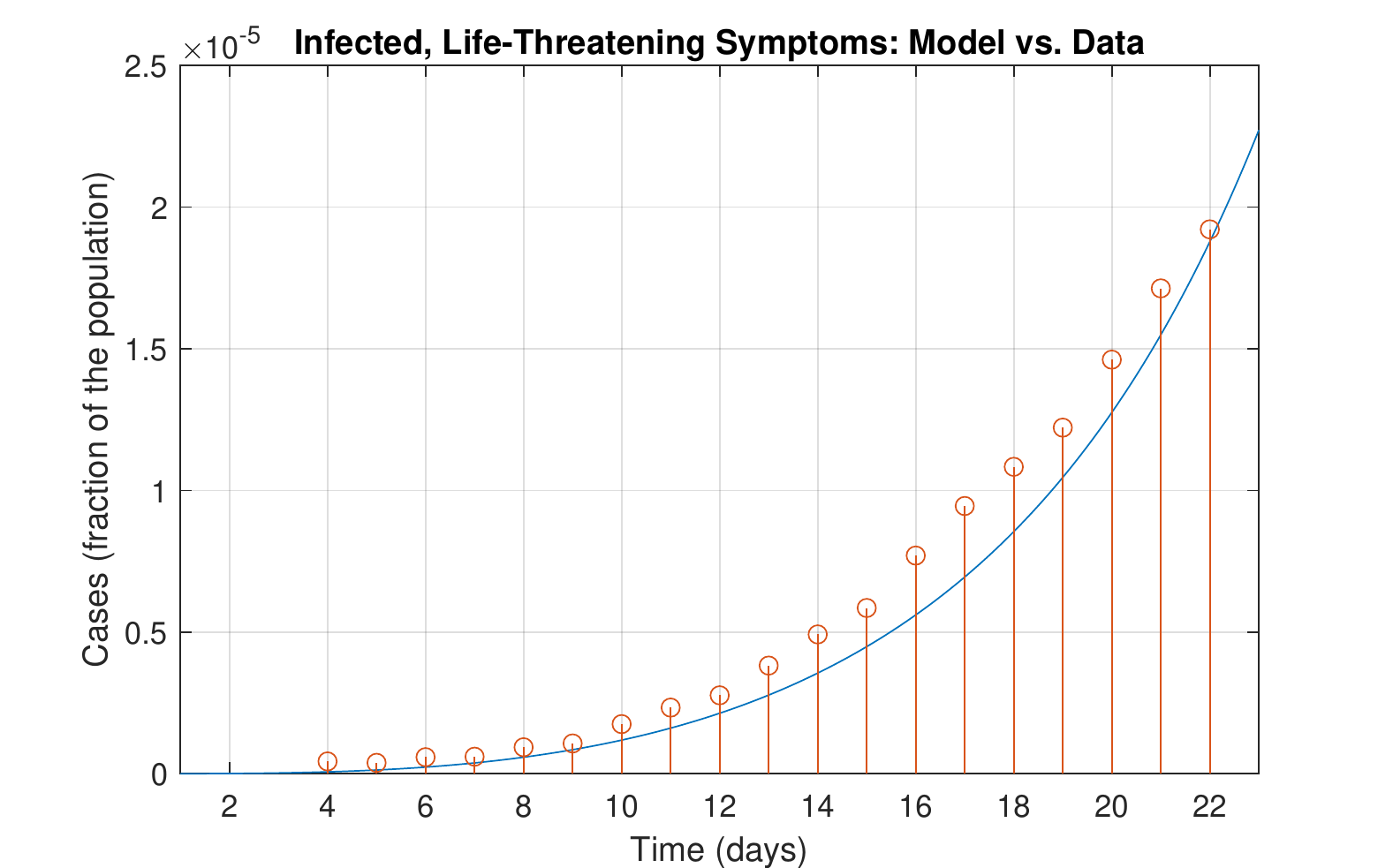}}
        \subfloat[Healed: $\int_0^t (\rho D(\phi)+ \xi R(\phi)+\sigma T(\phi))d\phi$.]{\includegraphics[width=.49\textwidth]{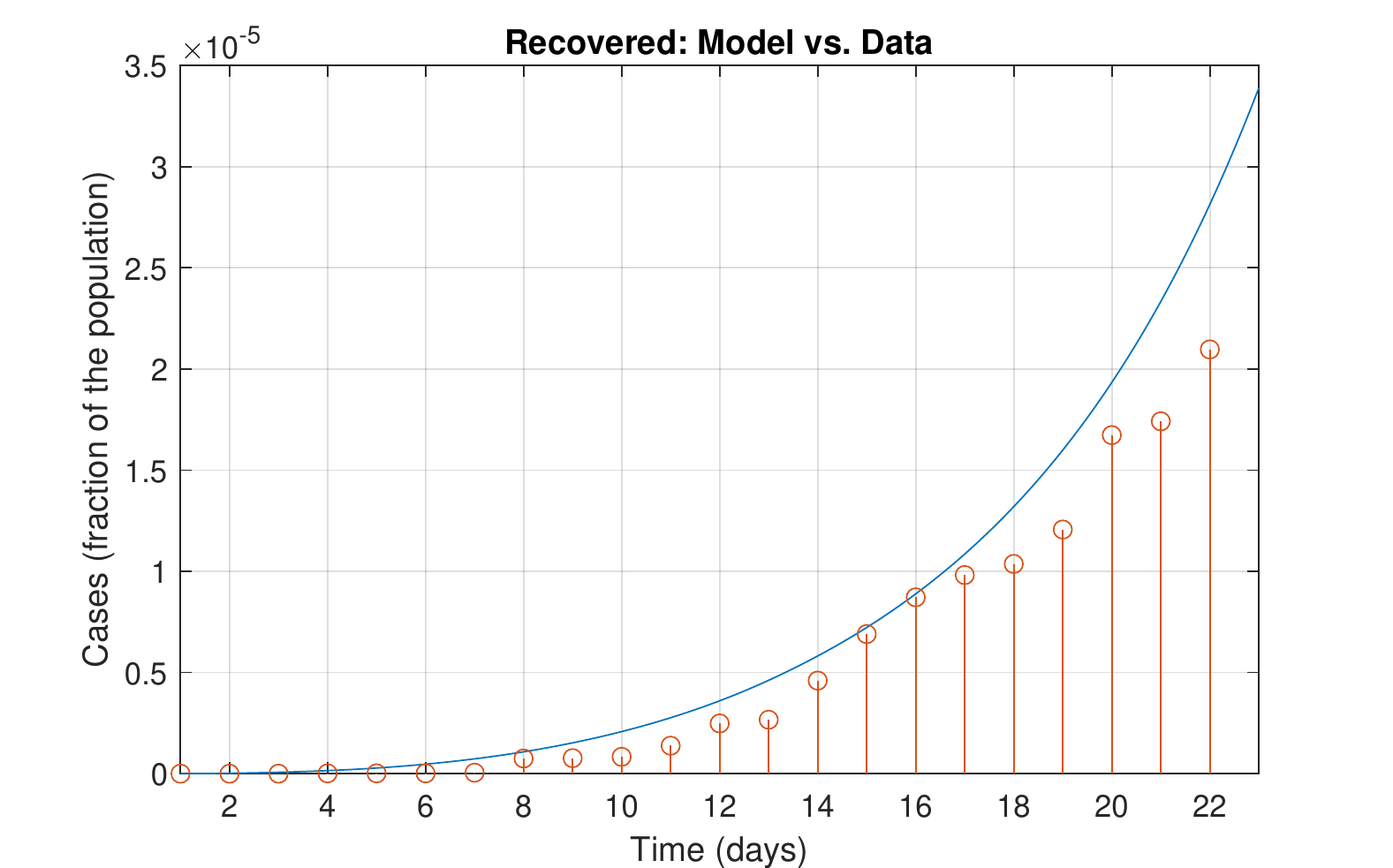}}\\
        \subfloat[Currently infected: $D(t)+R(t)+T(t)$.]{\includegraphics[width=.49\textwidth]{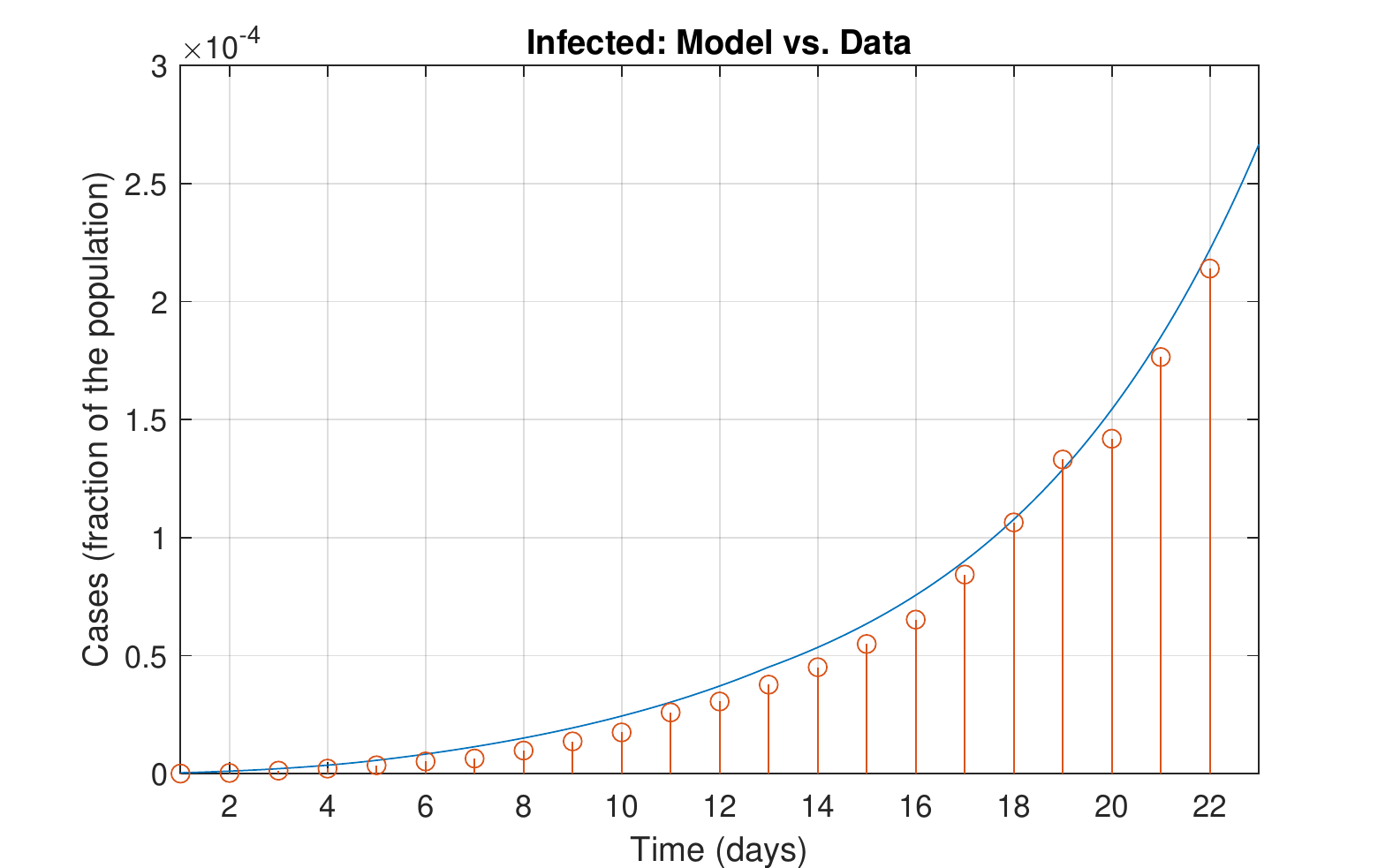}}
        \subfloat[Cumulative diagnosed cases: {$D(t)+R(t)+T(t)+E(t)+\int_0^t (\rho D(\phi)+ \xi R(\phi)+\sigma T(\phi))d\phi$.}]{\includegraphics[width=.49\textwidth]{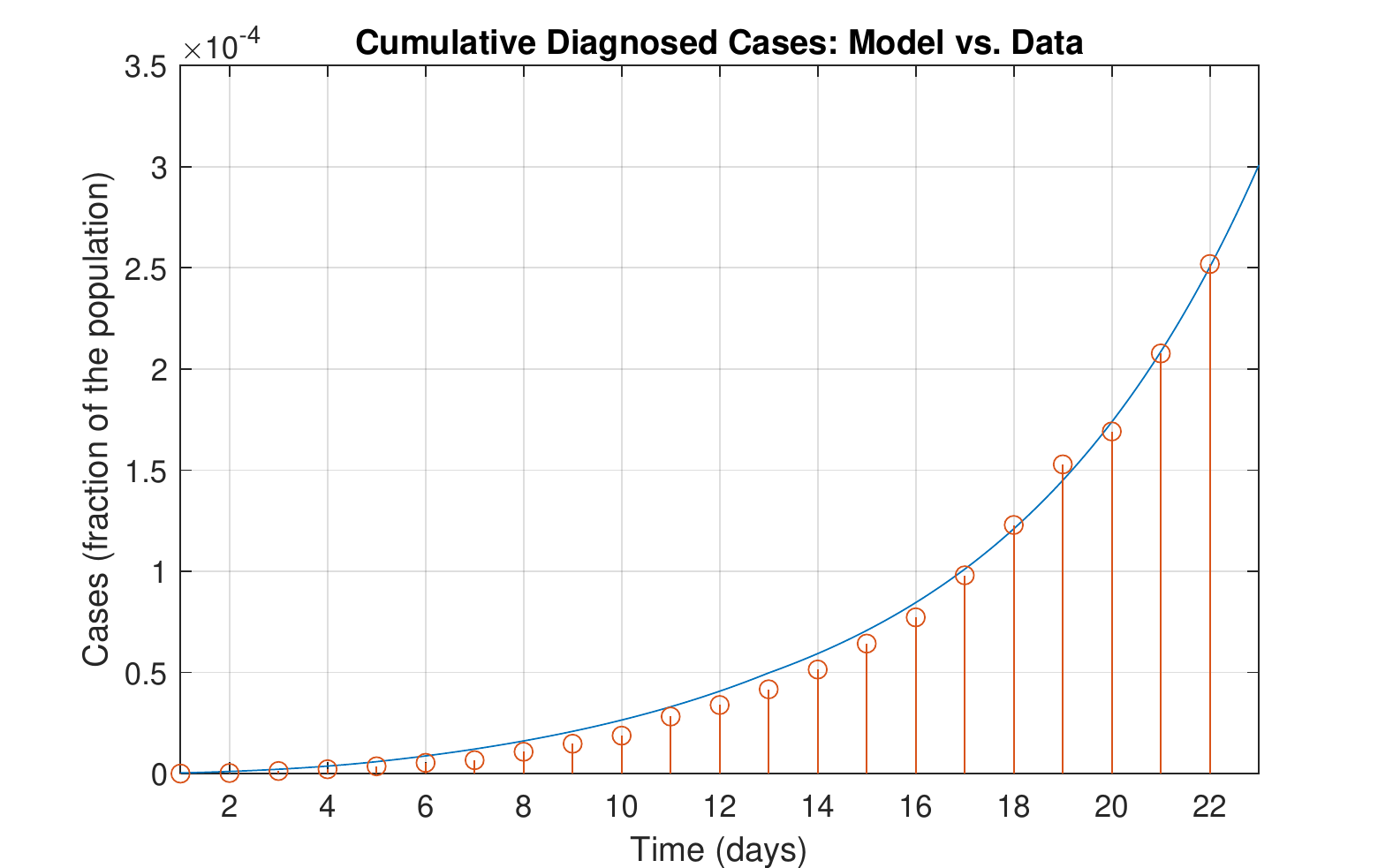}}
        \caption{Comparison between the official data (red dots histogram) and the results with the calibrated SIDARTHE model (blue line).}
        \label{fig:FIT}
\end{figure}

We infer the model parameters based on the official data (source: \emph{Protezione Civile} and \emph{Ministero della Salute}) about the evolution of the epidemic in Italy beginning February 20, 2020 (day $1$) through March 12, 2020 (day $22$). 
The data are turned into fractions over the whole Italian population (about $60$ millions).
The comparison between the official data and the curves resulting from the SIDARTHE model is reported in Figure \ref{fig:FIT}.

The estimated parameter values are based on the data about the number of currently infected individuals with different SOI (essentially asymptomatic, quarantined at home, corresponding to variable $D(t)$ in our model; symptomatic and hospitalised, corresponding to variable $R(t)$ in our model; symptomatic in life-threatening conditions, admitted to ICUs, corresponding to variable $T(t)$ in our model), and the number of diagnosed individuals who recovered (which can be computed as $\int_0^t [\rho D(\phi)+\xi R(\phi) + \sigma T(\phi)] d\phi$).

Data about the number of deaths (corresponding to $E(t)$ in our model) appear particularly high with respect to the case fatality rate reported in the literature; this can be largely explained by the age structure of the Italian population, which is the second oldest in the world (and the reported CFR across all countries is steeply increasing with the age of the patient) and by the extensive intergenerational contacts in the Italian society, which enhanced the spreading of the virus among older and more fragile generations \cite{Dowd2020}. Perhaps more importantly, it can be also explained by the Italian criteria for (provisional) statistics, which lead to overestimation. In fact, contrary to other countries, the official numbers for COVID-19 deaths \emph{provisionally} count the deaths of \emph{all the people tested positive for the SARS-CoV-2 virus}, even when they had multiple \emph{pre-existing life-threatening diseases} and the exact cause of death has not been ascertained yet, so these numbers \emph{still need to be confirmed} \cite{ISS}.

Thus, an important challenge in tuning the model is that the initial data are affected by statistical distortion: in particular, the values of the ratio death/infected is highly overestimated. The model fitting process must take this problem into account.
Therefore, we decide to fit the parameters based on the data about the diagnosed infected population and the number of recovered diagnosed patients, but not on the data about deaths.
It is also worth stressing that, in the long run, the model is weakly sensitive to the initial conditions; for this reason, the initial mismatch concerning the mortality data has little impact.

\section*{Acknowledgements}

We thank for their huge efforts the whole COVID19 IRCCS  San Matteo Pavia Task Force.\\
\emph{ID Staff}: Raffaele Bruno, Mario U Mondelli, Enrico Brunetti, Angela Di Matteo, Elena Seminari, Laura Maiocchi, Valentina Zuccaro, Layla Pagnucco, Bianca Mariani, Serena Ludovisi, Raffaella Lissandrin Aldo Parisi, Paolo Sacchi, Savino FA Patruno, Giuseppe Michelone, Roberto Gulminetti, Domenico Zanaboni, Stefano Novati, Renato Maserati, Paolo Orsolini, Marco Vecchia.
\emph{ID Residents}: Marco Sciarra, Erika Asperges, Marta Colaneri, Alessandro Di Filippo, Margherita Sambo, Simona Biscarini, Matteo Lupi, Silvia Roda, Teresa Chiara Pieri, Ilaria Gallazzi, Michele Sachs, Pietro Valsecchi.\\
\emph{Emergency Care Unit}: \emph{ECU Staff}: Stefano Perlini, Claudia Alfano, Marco Bonzano, Federica Briganti, Giuseppe Crescenzi, Anna Giulia Falchi, Roberta Guarnone, Barbara Guglielmana, Elena Maggi, Ilaria Martino, Pietro Pettenazza, Serena Pioli di Marco, Federica Quaglia, Anna Sabena, Francesco Salinaro, Francesco Speciale, Ilaria Zunino;
\emph{ECU Residents}: Marzia De Lorenzo, Gianmarco Secco, Lorenzo Dimitry, Giovanni Cappa, Igor Maisak, Benedetta Chiodi, Massimiliano Sciarrini, Bruno Barcella, Flavia Resta, Luca Moroni, Giulia Vezzoni, Lorenzo Scattaglia, Elisa Boscolo, Caterina Zattera, Michele Fidel Tassi, Vincenzo Capozza, Damiano Vignaroli, Marco Bazzini;
\emph{Intensive Care Unit}: Giorgio Iotti, Francesco Mojoli, Mirko Belliato, Luciano Perotti, Silvia Mongodi, Guido Tavazzi;
\emph{Paediatric Unit}: Gianluigi Marseglia, Amelia Licari, Ilaria Brambilla;
\emph{Virology Staff}: Daniela Barbarini, Antonella Bruno, Patrizia Cambieri, Giulia Campanini,  Giuditta Comolli, Marta Corbella, Rossana Daturi, Milena Furione, Bianca Mariani, Roberta Maserati, Enza Monzillo, Stefania Paolucci, Maurizio Parea, Elena Percivalle, Antonio Piralla, Francesca Rovida, Antonella Sarasini, Maurizio Zavattoni;
\emph{Virology Technical staff}: Guy Adzasehoun, Laura Bellotti, Ermanna Cabano, Giuliana Casali, Luca Dossena, Gabriella Frisco, Gabriella Garbagnoli, Alessia Girello, Viviana Landini, Claudia Lucchelli, Valentina Maliardi, Simona Pezzaia, Marta Premoli;
\emph{Virology Residents}: Alice Bonetti, Giacomo Caneva, Irene Cassaniti, Alfonso Corcione, Raffella Di Martino, Annapia Di Napoli, Alessandro Ferrari, Guglielmo Ferrari, Loretta Fiorina, Federica Giardina, Alessandra Mercato, Federica Novazzi, Giacomo Ratano, Beatrice Rossi, Irene Maria Sciabica, Monica Tallarita, Edoardo Vecchio Nepita;
\emph{Pharmacy Unit}: Monica Calvi, Michela Tizzoni.\\
\emph{Hospital Leadership}: Carlo Nicora, Antonio Triarico, Vincenzo Petronella, Carlo Marena, Alba Muzzi, Paolo Lago.\\
\emph{Data Unit}: Francesco Comandatore, Gherard Bissignandi, Stefano Gaiarsa, Marco Rettani, Claudio Bandi.

\section*{Author contributions}
G.G., F.B. and P.C. proposed the model and performed the mathematical derivations, the fitting and the simulations. R.B., A.D.F., A.D.M. and M.C. provided first-hand insight into the disease evolution and provided the clinical contextualisation and interpretation of the results. All authors wrote and approved the manuscript.

\section*{Competing interests}
The authors declare no competing interests.

\end{document}